\documentclass[aps,pre,reprint,showpacs,floatfix]{revtex4-1}
\usepackage{graphicx,amssymb,amsmath}
	
\usepackage{graphics}
\usepackage{natbib}

\DeclareMathOperator\erfc{erfc}

\begin{document}

\title{Effects of aging in catastrophe on the steady state and dynamics of a microtubule population}

\author{V. Jemseena}
\email{jemseena@physics.iitm.ac.in}
\author{Manoj Gopalakrishnan}
\email{manojgopal@iitm.ac.in (corresponding author)}

\affiliation{Department of Physics, Indian Institute of Technology Madras, Chennai 600036, India}

\date{\today}

\begin{abstract}
Several independent observations have suggested that catastrophe transition in microtubules is not a first-order process, as is usually assumed.  Recent {\it in vitro} observations by Gardner et al.[ M. K. Gardner et al., Cell {\bf147}, 1092  (2011)] showed that microtubule catastrophe takes place via multiple steps and the frequency increases with the age of the filament. Here, we investigate, via numerical simulations and mathematical calculations, some of the consequences of age dependence of catastrophe on the dynamics of microtubules as a function of the aging rate, for two different models of aging:  exponential growth, but saturating asymptotically and  purely linear growth. The boundary demarcating the steady state and non-steady state regimes in the dynamics is derived analytically in both cases. Numerical simulations, supported by analytical calculations in the linear model, show that aging leads to non-exponential length distributions in steady state. More importantly, oscillations ensue in 
microtubule length 
and 
velocity. The  regularity of oscillations, as characterized by the negative dip in the autocorrelation function, is reduced by increasing the frequency of rescue events. Our study shows that age dependence of catastrophe could function as an intrinsic mechanism to generate oscillatory dynamics in a microtubule population, distinct from hitherto identified ones. 
\end{abstract}


\maketitle

\section{Introduction}

The dynamic instability in microtubules  \cite{mitchison,desai}, in particular, the catastrophe transition by which a growing filament abruptly starts shrinking, has been the subject of extensive experimental and theoretical studies over several decades. The recently discovered phenomenon of  aging in microtubule catastrophe by Gardner et al. \cite{gardner} brings to light some new aspects of the catastrophe transition. Contradicting the long-standing view that (in the absence of  depolymerizing proteins) a filament is equally likely to undergo catastrophe at any instant of time, it was observed that the probability of a microtubule to undergo catastrophe  depends on how long it has been growing; the older the microtubule is,  the higher the probability to undergo  switching. Irrespective of tubulin concentration, the measured catastrophe frequency exhibited  a nonlinear dependence on the age of the microtubule, increasing linearly during  the early stage of growth and then approaching a steady state (
Fig. 3.E of Ref. \cite {gardner}). Also, the presence of depolymerizing proteins was observed to affect the aging process; for example, Kip3p accelerated the rate of aging while in the presence of MCAK, catastrophe was observed to become a first order process. Until this observation was made, it was generally believed that microtubule catastrophe is a single step stochastic event with no memory associated with it and hence usually modeled as a first order process. Within this frame work, many theoretical models have been set up in an attempt to understand the intrinsic dynamics of microtubules \cite{verde, dogterom, flyvbjerg,bicout}. It has been shown by Verde et al. \cite{verde}, the microtubule length distribution follows  a simple exponential decay in the bounded growth regime, when the switching rates are constants.  

Available experimental and simulation studies suggest that the origin of the aging phenomenon lies in the multi-filament structure of a microtubule \cite {gardner,gardner1}. A microtubule typically consists of 13-14 protofilaments wrapped into a cylindrical structure. During the early stages of growth, when all the protofilaments are of nearly equal length, the entire filament is structurally stable because of strong lateral bonding between protofilaments. As the filament grows longer, the tip of the microtubule becomes more and more tapered, weakening the lateral bonding between protofilaments and makes the filament more susceptible to undergo catastrophe. In this view, the "structural defects" responsible for the switching of the filament grows with time, until finally the filament undergoes catastrophe. An alternate view is that catastrophe requires the loss of guanosine triphosphate
(GTP) tip in a minimum number of protofilaments, and hence its kinetics could be visualized as a multi-step process, each step requiring the 
completion of the previous one for its materialization \cite{gardner,anderson,jemseena}. 
 
The possibility that the kinetics of plus end in microtubule dynamics may be non-first order was suggested first by Odde et al. nearly two decades ago \cite {odde}, and more recently by Stepanova et al. \cite{stepanova}. In  Ref. \cite{odde},  based on a probabilistic analysis of the experimentally observed growth time distributions, the authors showed that the plus and minus ends of a microtubule follow different kinetics. For the minus end, the growth time distribution could be described as exponential decay, characteristic of first order kinetics, whereas for the plus end, it turned out to be non-exponential; the experimental data in this case fitted well to gamma distribution.  Since the gamma distribution characterizes a multi step process, it was inferred that a growing microtubule (at the plus end) has to go through a sequence of events in order that catastrophe occurs. A new parameter was introduced to describe dynamic instability, apart from $\nu_c$ (catastrophe frequency), $\nu_r$ (rescue frequency)
, $v_g$ (growth velocity) and $v_s$ (shrinking velocity), viz., the  shape parameter $r$ of the growth time (or equivalently, time until catastrophe, hence also called catastrophe time) distribution (gamma distribution).  The value of the shape parameter that gave reasonably good fit with the experimental data is $r\sim3$  suggesting that catastrophe take place in a series of three steps, with each individual step being characterized by rate constant $\theta\sim1.67\mathrm{ min}^{-1}$. Later, Odde and Buettner \cite{odde1} showed that oscillations in state (growth/shrinkage) arise as a natural consequence of non-first order kinetics of catastrophe and rescue processes. 

\begin{figure}[h]
 \centering \includegraphics[height=5cm,width=7cm]{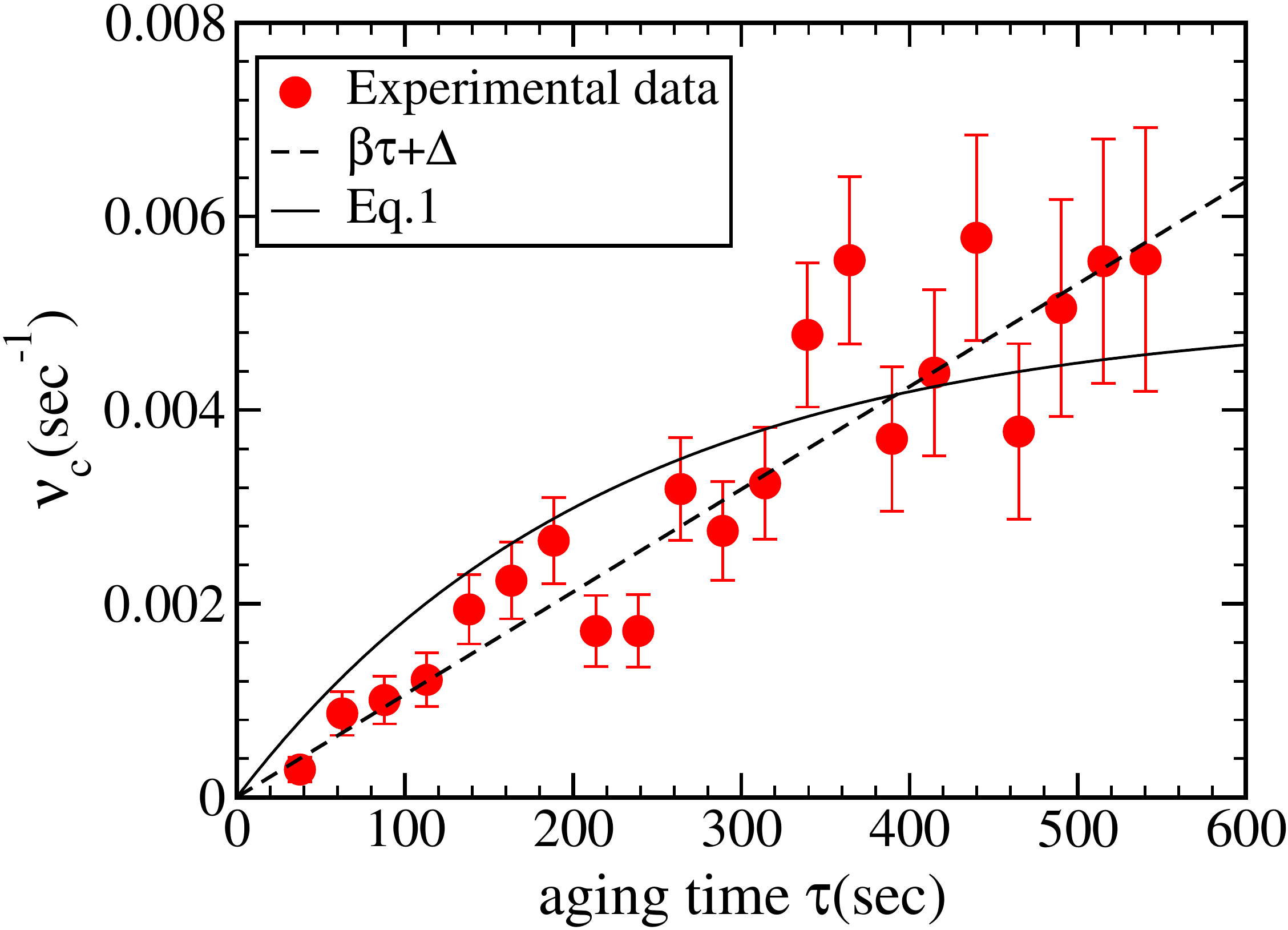}
 \caption{  A fit of the experimental data given in figure 1. D in Ref. \cite{gardner} , using the expression for age-dependent catastrophe frequency given by Eq.\ref{eq:eq1}. A simpler linear fit  given by Eq.\ref{eq:eqA4a} is also shown, see Sec. II for discussion. The best fit parameter values for the nonlinear curve given by Eq.\ref{eq:eq1} are $\nu_c^{\max}=0.3\mathrm {min}^{-1}$ and $\lambda=0.273\mathrm {min}^{-1}$, while for the linear curve $\beta=0.038\mathrm {min}^{-2}$ and $\Delta=0.0072\mathrm {min}^{-1}$.}
 \label{fig:fig0}
\end{figure}

In the present paper, we develop a mathematical model of dynamic instability, wherein the evolution of catastrophe frequency with the age of the filament  is explicitly taken into account. Motivated by experimental observations of Gardner et al.\cite{gardner}, we assume here that age-dependent catastrophe frequency is given by the phenomenological expression  

\begin{equation}
\nu_c(\tau)=\nu_c^{\max}\big(1-e^{-\lambda\tau}\big),
\label{eq:eq1}
\end{equation}
where $\nu_c^{\max}$ is the asymptotic value, $\lambda$ is referred to as the aging rate, with $\tau$ being the {\it age}, the time spent by the filament in the growing state after the last rescue event. In this sense, we assume the age of a filament as being given by an internal clock in the filament, the reading of which is reset to zero at every catastrophe, and which starts ``ticking'' as soon as rescue happens. For simplicity, and for lack of strong evidence to the contrary, we assume that rescue events follow first-order kinetics characterized by rate $\nu_r$. A fit of Eq.\ref{eq:eq1} to the experimental data in \cite{gardner} is shown in Fig. \ref{fig:fig0}. 

\vspace{0.5cm}

The motivation for choosing the above expression is primarily its mathematical simplicity, which makes it suitable for further calculations. A similar, but not exactly identical, mathematical form for aging of catastrophe also emerges naturally within a recently proposed model\cite{jemseena} where the  transition is modeled as a first passage process, such that a threshold number of protofilaments are required to lose their individual GTP tips by hydrolysis for it to happen. For details, we refer the reader to Supplementary Information \cite {supple}. In Ref. \cite{gardner}, by comparison, it is assumed that the catastrophe time distribution (to be discussed in Sec. II) is represented by gamma distribution, similar to \cite{odde}. Here too, the catastrophe transition is visualized as a first passage event, and requires the occurrence of a threshold number $r$ of underlying events, but each event is assumed to occur with the {\it same} rate $\theta$. In this case, the precise nature of the underlying 
events is not 
specified in detail. Both the exponential aging function in Eq.\ref{eq:eq1} as well as the corresponding function derived from the gamma distribution fit the experimental data almost equally well; see Fig. 1 in Supplemental Material \cite{supple}.

A brief summary of our results are as follows: as a direct consequence of Eq.\ref{eq:eq1}, the distribution of time intervals corresponding to the growth phase is found to be non-monotonic (Section II). The Fokker-Planck equations for microtubule dynamics in the presence of aging (set up  in Sec. III) lead to a number of analytical results, presented in Sec. IV. In particular, the stationary length distribution is non-exponential, approaching half-Gaussian form in the limit $\lambda\to 0$, while the standard exponential decay is recovered in the limit $\lambda\to\infty$. Exact mathematical expressions for length autocorrelation function are derived in Laplace space, but in general, explicit inversion is found difficult. Numerical simulations (Sec. IV) show that the autocorrelation function possesses a negative lobe, signifying oscillatory behavior, the half-period of which is measured as a function of aging rate $\lambda,\nu_c^{\max}$ and $v_g/v_s$, the ratio of growth and shrinkage velocities. 
Numerically computed velocity autocorrelation function also shows similar properties.  
For near-experimental parameter values, the half-period of length oscillations is found to be in the range of $\sim 5-10$ minutes. In this context, we also make a comparison with experimental observations on chromosome oscillations in the concluding section (Sec. V). 

\section{Catastrophe time distribution}
   
In this paper, we introduce the age of the filament $\tau$  as a variable that describes the state of a growing filament (in addition to its length) irrespective of the origins of the aging effect. Although rescue events are not reported in the experiments by Gardner et al. \cite{gardner}, here we take a step a forward: we assume that if rescue events are present, after every rescue, a filament starts to grow as a fresh filament without any structural defects.  Age is reset to zero at each catastrophe event, which again starts to grow linearly with time upon  the rescue of the filament. Since our objective is to explore the consequences of age-dependent catastrophe on dynamic instability, we do not consider explicitly microscopic events such as addition/removal of monomers and hydrolysis; these enter the formalism implicitly through catastrophe and rescue rates\cite{flyvbjerg,antal,ranjith,jemseena}. 

In the present study, we use the two state continuum approach model proposed by Verde et al. \cite{verde}, where the dynamics is captured by  a pair of coupled first order partial differential equations at a macroscopic level.  Many mathematical models have been formulated in the past, based on this two state approach to explore dynamic instability and its applications, eg., microtubule oscillations \cite {vallade, flyvbjerg1}, dynamics under confinement\cite{bindu,mulder}, search for targets\cite{holy,manoj,mulder} and the effects of regulators \cite{sutapa,tischer}.
 
\vspace{0.1cm}
\begin{figure}[h]
 \centering \includegraphics[height=5cm,width=6.5cm]{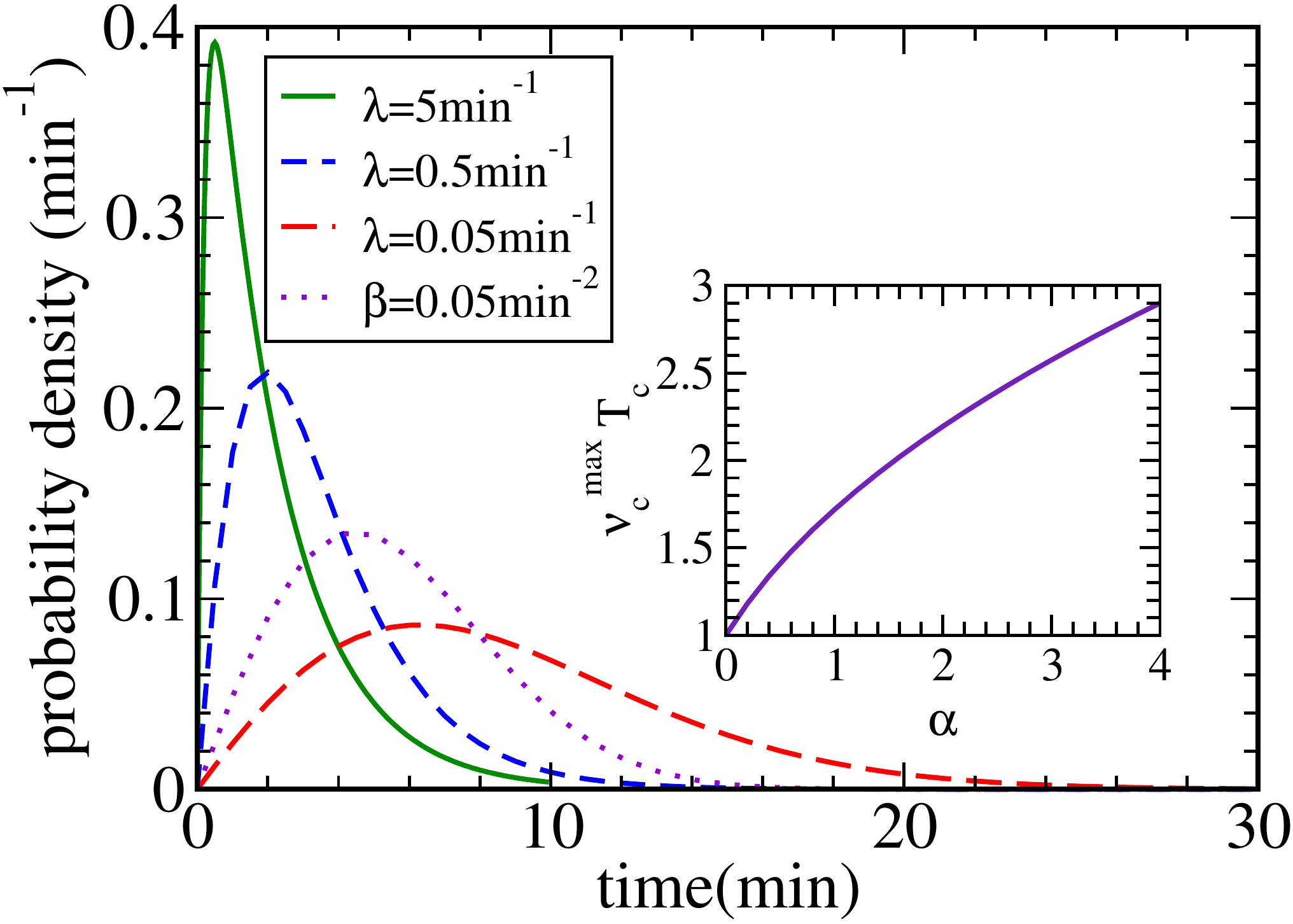}
 \caption{Probability distribution function for catastrophe time (growth time) given by Eq. \ref{eq:eqA2},  for different $\lambda$ values with $\nu_c^{\max}=0.5\mathrm {min}^{-1}$. The dotted curve gives the catastrophe time distribution for the linear model with $\beta=0.05 \mathrm {min}^{-2}$. In the inset the scaled average catastrophe time for the exponential model, given by Eq. \ref{eq:eqA6} as a function of $\alpha$ is shown.}
 \label{fig:fig1}
\end{figure}

Let us denote by $T$ the time until catastrophe (referred to as the {\it catastrophe time}), starting from a rescue event at $T=0$. Then, the probability distribution for $T$ is given by 

\begin{equation}
f_c(T)=\nu_c(T)\exp\left(-\int_0^T \nu_c(\tau)d\tau\right).
\label{eq:eqA1}
\end{equation}
After substituting Eq.\ref{eq:eq1} in Eq.\ref{eq:eqA1}, we find 
\begin{equation}
f_c(T)= \nu_c^{\max}  e^{\alpha}\bigg[ (1-e^{-\lambda T}) \exp[-(\nu_c^{\max}T+\alpha e^{-\lambda T} )]\bigg],
\label{eq:eqA2}
\end{equation}
for which the following limiting cases are of special interest: 

\begin{eqnarray}
f_c(T)\sim \nu_c^{\max} e^{\alpha}\exp[-\nu_c^{\max}T],~~~~\lambda T\gg 1\\
f_c(T)\sim \beta T e^{\frac{-\beta T^2}{2}},~~~~~~~~~~~~\lambda T\ll 1,~
\label{eq:eqA3}
\end{eqnarray}
where we have defined the parameters
\begin{equation}
\alpha=\frac{\nu_c^{\max}}{\lambda}~~;~~\beta=\nu_c^{\max}\lambda. 
\label{eq:eqA4}
\end{equation}

Note that the latter regime ( $\lambda  T \ll 1$), if extended to all times,  formally corresponds to a linear dependence of catastrophe frequency on age, i.e., 
\begin{equation}
 \nu_c(\tau)=\beta\tau,
 \label{eq:eqA4a}
\end{equation}
which we shall call the ``linear model", to distinguish it from the ``exponential model" given by Eq.\ref{eq:eq1} (note that even the linear model fits the experimental data reasonably well, see Fig.\ref{fig:fig0}). Both models will be studied in this paper. The linear model turns out to be more amenable to mathematical analysis; see, for instance, Sec. IV B for explicit analytical expression for length  autocorrelation function. 

The distributions in Eq.\ref{eq:eqA2} and Eq.\ref{eq:eqA3} are non-monotonic function of $T$ (see Fig.\ref{fig:fig1}), with the maxima, corresponding to the most probable switching time, located at

\begin{eqnarray}
T_{\mathrm m}= \frac{1}{\lambda}  \ln\bigg[ \frac{2\alpha}{(1+2\alpha-\sqrt{1+4\alpha})}\bigg] ~~\mathrm {(exponential~ model)},\nonumber\\
T_{\mathrm m}=\frac{1}{\sqrt{\beta}}~~~~~~~~~~~~~~~\mathrm {(linear~ model)}.~~~~~~~~~~~~~~~~~
\label{eq:eqA4+}
\end{eqnarray}

In the limit $\alpha \gg 1$, the expression for the exponential model approaches that of the linear model, as expected. The average time until catastrophe is calculated as 

\begin{equation}
T_c=\int dT f_c(T)T.
\label{eq:eqA5}
\end{equation}

Using Eq.\ref{eq:eqA2} and Eq.\ref{eq:eqA5}, we find, for the exponential model,

\begin{equation}
T_c= \frac{\alpha ^2 e^{\alpha}}{\nu_c^{\max}}\sum_{n=0}^{\infty}\frac{(-1)^n \alpha ^n}{n!}\bigg[\frac{1}{(\alpha+n)^2}-  \frac{1}{(\alpha+(n+1))^2}\bigg],
\label{eq:eqA6}
\end{equation}
with the  limiting behavior $T_c\sim 1/\nu_c^{\max}$ as $\lambda\rightarrow\infty$, corresponding to the constant catastrophe case. For the linear model in Eq.\ref{eq:eqA4a}, we find, similarly, 

\begin{eqnarray}
T_c=\frac{1}{4}\sqrt{\frac{\pi}{\beta}}.
\label{eqA7}
\end{eqnarray}

The non-monotonic behavior of $f_c(T)$, by virtue of the ``tightening" of the direction reversal times (see Fig. \ref{fig:fig1}), opens up the possibility of oscillatory dynamics, which we shall explore in detail in Sec. IV B later using  autocorrelation functions. Numerical simulations indicate that oscillations occur when the aging rate (in the exponential model) and rescue frequency are sufficiently small. 
 
\section{Continuum equations for the two-state stochastic model}

According to the two-state stochastic model, a given microtubule exists either in the growing state or shrinking state during its life time. A third state called pause state is also possible where the filament neither grows or shrinks \cite{desai}. In the present study, we ignore the existence of the pause state, as well as the structural details of the filament, although such details may implicitly affect the catastrophe rate.  A microtubule nucleates from a nucleation center at a rate $\nu$ (`birth') and it may disappear completely by shrinking to length zero (`death'). In between birth and death, a microtubule switches between growing and shrinking states; switch from  growing state to shrinking state (catastrophe) takes place at rate $\nu_c(\tau)$ (given by Eq.\ref{eq:eq1}, including the limiting linear form) and the reverse transition takes place at rate $\nu_r$ (rescue) which is taken to be a constant. The growth velocity is denoted by $v_g$, the shrinkage velocity by $v_s$, and the length by $x$. 

Let us now denote by $G_{1j}^{\prime}(x,\tau,t|x_0,0)dxd\tau$ the probability to find a filament in the growing state with length and age in the range $[x,x+dx]$ and $[\tau,\tau+d\tau]$ respectively, at time $t$. Similarly, $G_{0j}(x,t|x_0)dx$ gives the probability to find a filament in the shrinking state with length in the range $[x,x+dx]$ at time $t$. In both cases, $x_0$ is the initial length and $j=1,0$, indicates the state of growth or shrinkage at $t=0$. Note that the age variable $\tau$ is relevant only in the growing state, and is therefore present only in one set of Green's functions, for which the primed notation is used. 

The variables $x$ and $\tau$ evolve according to the following equations:

\begin{eqnarray}
\frac{dx}{dt}=v_g;~~~~\frac{d\tau}{dt}=1~~~~~  (\mathrm{growing ~ state}),~~ \nonumber \\
\frac{dx}{dt}=-v_s;~~~~~\tau=0~~~~(\mathrm{shrinking ~ state}).
\label{eq:eqB2-}
\end{eqnarray}

The Green's functions then satisfy the following differential equations, which suitably generalize the corresponding equations for constant catastrophe rate, with aging absent\cite{verde, dogterom,bicout}: 

\begin{widetext}
  \begin{eqnarray}
\frac{\partial G_{1j}^{\prime}(x,\tau,t|x_0,0)}{\partial t}=-v_g \frac{\partial G_{1j}^{\prime}(x,\tau,t|x_0,0)}{\partial x}-\frac{\partial G_{1j}^{\prime}(x,\tau,t|x_0,0)}{\partial \tau}  -\nu_c(\tau) G_{1j}^{\prime}(x,\tau,t|x_0,0)+\nu_r G_{0j}(x,t|x_0) \delta(\tau),
\label{eq:eqB1}
\end{eqnarray}

\begin{eqnarray}
\frac{\partial G_{0j}(x,t|x_0)}{\partial t}=v_s  \frac{\partial G_{0j}(x,t|x_0)}{\partial x}+\int\limits_{0}^{\infty} d\tau~ \nu_c(\tau) G_{1j}^{\prime}(x,\tau,t|x_0,0) -\nu_r G_{0j}(x,t|x_0), ~~~~~~~~~~~~~~~~x\geq0.~~~~~~~~
\label{eq:eqB2}
\end{eqnarray}
\end{widetext}

Note that resetting of age to zero in the shrinking state is achieved through the introduction of delta function in Eq.\ref{eq:eqB1}.The Green's functions satisfy the normalization condition: $\int\limits_{0}^{\infty} [ G_{1j}(x,t|x_0) +G_{0j}(x,t|x_0)] dx=1$, where 
\begin{equation}
G_{1j}(x,t|x_0)\equiv \int\limits_{0}^{\infty} G_{1j}^{\prime}(x,t,\tau|x_0,0) d\tau.
\label{eq:eqB2+}
\end{equation}
 
\section{Results}
\subsection{Length and age distribution}

Many microtubule-dependent functions in the cell depend on the spatial organization of microtubules, and hence, a quantity of primary interest is their length distribution. In the case of constant switching rates, Verde et al. \cite{verde} have shown that a microtubule population reaches a steady state with a stationary length distribution, when 
\begin{equation}
\nu_c v_s> \nu_r v_g,
\label{eq:eqB2+}
\end{equation}
in which case the lengths are exponentially distributed, with mean length $\langle x\rangle=(\nu_c/v_g-\nu_r/v_s)^{-1}$. To understand how these results are modified in the presence of age-dependence in catastrophes, we start by defining the stationary length distributions
\begin{eqnarray}
P_1^{\prime}(x,\tau)\equiv \underset{t \to \infty}{\lim} G_{1j}^{\prime}(x,\tau,t|x_0,0),\nonumber \\ 
P_0(x)\equiv \underset{t \to \infty}{\lim} G_{0j}(x,t|x_0),~~~~~~~
\label{eq:eqC1}
\end{eqnarray}
which satisfy the following steady state equations, obtained by putting the LHS to zero in Eq.\ref{eq:eqB1} and Eq.\ref{eq:eqB2}.

\begin{eqnarray}
v_g \frac{\partial P_1^{\prime}(x,\tau)}{\partial x}= -\frac{\partial P_1^{\prime}(x,\tau)}{\partial \tau}-\nu_c(\tau) P_1^{\prime}(x,\tau),\nonumber \\ +\nu_r P_0(x) \delta(\tau)~~~~~~~~~~~~~~
\label{eq:eqC2}
\end{eqnarray}

\begin{eqnarray}
v_s \frac{\partial P_0(x)}{\partial x}= -\int \limits_{0}^{\infty} \nu_c(\tau) P_1^{\prime}(x,\tau) d\tau +\nu_r P_0(x).~~~~
\label{eq:eqC3}
\end{eqnarray}

To solve Eq.\ref{eq:eqC2} and Eq.\ref{eq:eqC3} we assume, for simplicity, that nucleation rate is very large such that nucleation of a new filament takes place instantaneously once a microtubule disappears at that site; hence the net flux at the boundary is zero.  The resulting flux balance condition at the boundary $x=0$ can be written as 

\begin{eqnarray}
 v_s P_0(x) \delta(\tau) \vert_{x=0} =v_g P_1^{\prime}(x,\tau)\vert_{x=0} 
 \label{eq:eqC4}
\end{eqnarray}
 and we define, $J_0\equiv v_s P_0(x)\vert_{x=0}$.
We now Laplace-transform $x\to p$ in Eq.\ref{eq:eqC2} and Eq.\ref{eq:eqC3} by defining
\begin{eqnarray}
\tilde P_1^{\prime}(p,\tau)=\int \limits_{0}^{\infty} e^{-px} P_1^{\prime}(x,\tau) dx; \tilde P_0(p)=\int \limits_{0}^{\infty} e^{-px} P_0(x) dx.\nonumber\\
\label{eq:eqC5}
\end{eqnarray}

After applying the above transformations in Eq.\ref{eq:eqC2} and Eq.\ref{eq:eqC3}, followed by simplifications (see Appendix A for details), we arrive at the expression

\begin{eqnarray}
\tilde P_1^{\prime}(p,\tau)= \bigg[\frac{J_0 v_sp }{v_s p-\nu_r(1-\zeta(p))} \bigg] \times~~~~~~~~~~~~~~  \nonumber \\ \exp\bigg( -v_g p \tau -\int \limits_{0}^{\tau} \nu_c(\tau^{\prime}) d\tau^{\prime} \bigg),
\label{eq:eqC6}
\end{eqnarray}
 where the function $\zeta(p)$ is given by Eq.\ref{eq:eqAA8}. After integrating out $\tau$ from Eq.\ref{eq:eqC6}, we obtain

\begin{eqnarray}
 \tilde P_1(p)= \bigg[\frac{J_0 v_sp }{v_s p-\nu_r(1-\zeta(p))} \bigg] \eta(p), 
\label{eq:eqC7}
\end{eqnarray}
$\eta(p)$ is  given by Eq.\ref{eq:eqAA11}.
 Similarly,

\begin{eqnarray}
\tilde P_0(p)=\frac{J_0 (1-\zeta(p)) }{v_s p-\nu_r(1-\zeta(p))}. ~~~~~~~~~~~~
\label{eq:eqC8}
\end{eqnarray}

 Both Eq.\ref{eq:eqC7} and Eq.\ref{eq:eqC8}, together with the normalization property of the probability distribution functions
(PDFs) are used to find the  expression for $J_0$ for specific forms of $\nu_c(\tau)$. \\


{\it Exponential aging: }We use the form of $\nu_c(\tau)$ given in Eq.\ref{eq:eq1} to calculate $\zeta(p)$ and $\eta(p)$  from the integrals given by  Eq.\ref{eq:eqAA8} and Eq.\ref{eq:eqAA11}. The expression for $J_0$, fixed using normalization, i.e., $\sum_{j=0,1}{\tilde P}_j(0)=1$, turns out to be 

\begin{eqnarray}
J_0=\frac{\nu_c^{\max} v_s-\nu_r v_g [1+e^{\alpha}\alpha^{-\alpha} \gamma(\alpha+1,\alpha)]}{(v_g+v_s)[1+ e^{\alpha}\alpha^{-\alpha} \gamma(\alpha+1,\alpha)]}.
\label{eq:eqC9}
\end{eqnarray}

 Here $\gamma(\alpha+1,\alpha)$ is the lower incomplete gamma function, defined \cite{abramowitz}
 \begin{equation}
  \gamma(a,z)=\int \limits_{0}^{z} e^{-t}t^{a-1} dt, 
  \label{eq:eqC10}
 \end{equation}
 and the dimensionless constant $\alpha$ is defined in Eq.\ref{eq:eqA4}.  In the limit $\alpha\rightarrow0$ ($\lambda\rightarrow\infty$), the term in the square bracket approaches unity. Steady state is guaranteed only if the numerator of the expression in Eq.\ref{eq:eqC9} is positive, thus for age-dependent catastrophe of the form in Eq.\ref{eq:eq1}, the condition given by Verde et al.  \cite{verde} is modified as 

\begin{equation}
\nu_c^{\max} v_s>\nu_r v_g [1+e^{\alpha}\alpha^{-\alpha} \gamma(\alpha+1,\alpha)]. 
\label{eq:eqC11}
\end{equation}

In the limit $\lambda\rightarrow\infty$ the above condition  reduces to $\nu_c^{\max} v_s>\nu_r v_g$. 

Unfortunately, it is not possible to invert the Laplace-transforms in Eq.\ref{eq:eqC7} and Eq.\ref{eq:eqC8} explicitly, except for the special (but experimentally relevant) case  $\nu_r=0$. After substitution of the required $\zeta(p)$ and $\eta(p)$ from Eq.\ref{eq:eqAA16} and Eq.\ref{eq:eqAA17} respectively, we find that the length distribution $P(x)=\sum_{j=0,1} P_j(x)$ has the form (for details see Appendix B)

\begin{eqnarray}
P(x)=\mathcal{A}\exp\left(-\frac{\nu_c^{\max}}{v_g}x -\alpha e^{ \frac {-\lambda}{v_g}x}\right),
 \label{eq:eqC12}
\end{eqnarray}
with the normalization constant given by 
\begin{eqnarray}
\mathcal{A} =\frac{\nu_c^{\max}e^{\alpha} } { v_g \bigg(1+ e^{\alpha}\alpha^{-\alpha} \gamma(\alpha+1,\alpha) \bigg)}.
 \label{eq:eqC13}
\end{eqnarray}

The state-specific distributions are given by $P_1(x)=\phi P(x)$ and $P_0(x)=(1-\phi)P(x)$ where $\phi=v_g/(v_g+v_s)$. The various limiting behaviours of the distribution in Eq.\ref{eq:eqC12} are as follows:

\begin{eqnarray}
P(x)\propto \exp\left(-\frac{\lambda\nu_c^{\max}}{2v_g^2}x^2\right),~~~\lambda\ll\nu_c^{\max}~\nonumber\\
P(x)\propto \exp\left[-\left(\frac{\nu_c^{\max}}{v_g}x+e^{-\frac{\nu_c^{\max}}{v_g}x}\right)\right],~~~\lambda=\nu_c^{\max}~\nonumber\\
P(x)\propto \exp\left(-\frac{\nu_c^{\max}}{v_g}x\right),~~~\lambda\gg\nu_c^{\max}.~~
\label{eq:eqC14}
\end{eqnarray}

The first two moments of the distribution in Eq.\ref{eq:eqC12} are of interest: these are 

\begin{equation}
\langle x\rangle=\frac{\mathcal{A} v_g^2}{\lambda^2}\sum_{n=0}^{\infty}\frac{(-1)^n \alpha ^n}{n!}\bigg[\frac{1}{(\alpha+n)^2}\bigg]
\label{eq:eqC15}
\end{equation}
 and
\begin{equation}
\langle x^2\rangle=\frac{2\mathcal{A} v_g^3}{\lambda^3}\sum_{n=0}^{\infty}\frac{(-1)^n \alpha ^n}{n!}\bigg[\frac{1}{(\alpha+n)^3}\bigg].
\label{eq::eqC16}
\end{equation}

{\it Linear aging }: It is also interesting to study separately the implications of linear increase in catastrophe with age. 
In this case, the current at the boundary turns out to be 
\begin{equation}
J_0=\frac{v_s-\nu_rv_g\sqrt{\frac{\pi}{2\beta}}}{(v_s+v_g) \sqrt{\frac{\pi}{2\beta}}}.
\label{eq:eqC17}
\end{equation}

The inverse Laplace transform of Eq.\ref{eq:eqC7} and Eq.\ref{eq:eqC8}, with $\eta(p)$ and $\zeta(p)$ given by expressions Eq.\ref{eq:eqAA23} and Eq.\ref{eq:eqAA24} respectively in Appendix B, leads to the length  distribution (in this case, half-Gaussian)
 
\begin{eqnarray}
 P(x)=\sqrt{\frac{2\beta}{\pi v_g^2}}~e^{-\frac{\beta}{2v_g^2}x^2},
 \label{eq:eqC18}
\end{eqnarray} 
which is identical to the first limiting case in Eq.\ref{eq:eqC14}.

Some parallels with the related problem of length-dependent catastrophe are worth mentioning here. In the absence of rescue, length of a filament is given by $v_g\tau$, i.e,  $x\propto\tau$, suggesting that in this case,  the age-dependent and length-dependent catastrophes are {\it effectively} the same, although the underlying mechanisms are quite different. Experimental evidence suggests that the intrinsic dynamics of microtubules can be regulated  by the presence of microtubule associated proteins.  A special class of proteins known as depolymerizing proteins are capable of enhancing the catastrophe rate. A widely studied example is the microtubule depolymerizer Kip3p, belonging to the kinesin 8 family, which is known to increase the catastrophe rate in a length-dependent manner \cite{varga}. Several authors have studied deploymerizer-induced length-dependent depolymerization and catastrophe \cite{tischer,manoj1,klein,frey,betterton}. Tischer et al. \cite{tischer} have shown that in the presence of 
length-dependent transition rates (both rescue and 
catastrophe) the length distribution develops a peak, implying a tighter regulation of lengths. In fission yeast experiments reported in Tischer et al.\cite{tischer1} where rescue was found to be absent but catastrophe increases nearly linearly with length of the filament, it was predicted that length distribution is half-Gaussian, identical to Eq.\ref{eq:eqC18}.

{\it Age distribution}: The steady state age distribution in the growing state is defined as 

\begin{eqnarray}
\rho(\tau)=\int \limits_{0}^{\infty} P_1^{\prime}(x,\tau) dx,
 \label{eq:eqC19}
\end{eqnarray}
where $P_1^{\prime}(x,\tau)$ is given by inverse Laplace transform of  Eq.\ref{eq:eqC6} with respect to $p$. The result (which is  independent of $\nu_r$), is 

\begin{eqnarray}
\rho(\tau)=\frac{\lambda \alpha^{\alpha}\exp(-\nu_c^{\max} \tau-\alpha e^{-\lambda \tau})}{\gamma(\alpha,\alpha)}~~(\mathrm {exponential ~model}) \nonumber \\
 \label{eq:eqC20}
\end{eqnarray}
with limiting behaviors

\begin{eqnarray}
\rho(\tau)\propto e^{-\nu_c^{\max}\tau},~~~\lambda\tau\gg 1\nonumber\\
\rho(\tau)\propto e^{-\frac{\beta\tau^2}{2}},~~~\lambda\tau\ll 1.
\label{eq:eqC20+}
\end{eqnarray}

In the linear model, the age distribution is half-Gaussian at all times, consistent with Eq.\ref{eq:eqC18}: 

\begin{eqnarray}
\rho(\tau)=\sqrt{\frac {2\beta}{\pi}}e^{-\frac{\beta\tau^2}{2}}~~~~(\mathrm{linear~model}).
\label{eq:eqC21}
\end{eqnarray}

\subsection{Autocorrelation functions}

We will now investigate further on the effects of aging, and the consequent change in the catastrophe time distribution, on correlation functions. In general, we expect that the observed peak in the distribution of catastrophe time (Fig.\ref{fig:fig1}) would make the catastrophe transition more deterministic and predictable, and hence may confer an oscillatory nature to microtubule dynamics. Microtubule oscillations is a well-studied phenomenon experimentally ~\cite{carlier, carlier1,mandelkow,mandelkow1,mandelkow2}, but here, the oscillations generally refer to periodic changes in free tubulin concentration due to polymerization and depolymerization of filaments. The exception is the work by Odde and Buettner\cite{odde1}; here, the authors considered dynamic instability as a two-state jump process with gamma-distributed residence times in each state. The autocorrelation function for the growth state was shown to exhibit oscillations.

In this study, we first looked at the autocorrelation in length of a microtubule assembly. In terms of the Green's functions $G_{ij}$ (Eq.\ref{eq:eqB1} and Eq.\ref{eq:eqB2}) and probability distributions $P_j$ defined in Sec. III and Sec. IV.A respectively, the partial autocorrelation functions for length is defined as 

\begin{eqnarray}
\langle x(0) x(t)\rangle_{ij}=\int dx_0\int dx x_0~x~ G_{ij}(x,t|x_0)P_j(x_0).~~
 \label{eq:eqC22}
\end{eqnarray}

Next, by summing Eq.\ref{eq:eqC22} over all the initial and final states, we define 

\begin{eqnarray}
 \mathcal{C}_{x}^{\prime}(t)=\sum\limits_{i,j}\langle x(0) x(t)\rangle_{ij},
  \label{eq:eqC23}
\end{eqnarray}
in terms of which the normalized autocorrelation is given by 

\begin{eqnarray}
 \mathcal{C}_{x}(t)=\frac {\mathcal{C}_{x}^{\prime}(t)-\langle x\rangle^2}{\sigma_x^2},
  \label{eq:eqC24}
\end{eqnarray}
where $\sigma_x^2=\langle x^2\rangle-\langle x\rangle^2$ is the variance in length. The normalization ensures that $C_x(0)=1$.

The calculations of the autocorrelation function for the exponential model turned out to be cumbersome, and hence explicit results were derived only in the limit $\lambda\to\infty$ (constant catastrophe). For the linear model, explicit results could be
obtained in Laplace space. We present our results for these cases now.

\subsubsection{Case 1: Constant catastrophe, $\lambda\rightarrow\infty$}

When we put $\lambda=\infty$ strictly, the Green's functions $G_{1j}$ depends only on $x$, and not on $\tau$. Then, Eq.\ref{eq:eqB1} and Eq.\ref{eq:eqB2} reduce to

\begin{eqnarray}
 \frac{\partial G_{1j}(x,t|x_0)}{\partial t}=-v_g \frac{\partial G_{1j}(x,t|x_0)}{\partial x}-\nu_c^{\max} G_{1j}(x,t|x_0)\nonumber \\ +\nu_r G_{0j}(x,t|x_0),~~~~~~~~~~~~~~~~~~~
  \label{eq:eqE1}
\end{eqnarray}
\begin{eqnarray}
 \frac{\partial G_{0j}(x,t|x_0)}{\partial t}=v_s \frac{\partial G_{0j}(x,t|x_0)}{\partial x}+\nu_c^{\max} G_{1j}(x,t|x_0) \nonumber \\ -\nu_r G_{0j}(x,t|x_0).~~~~~~~~~~~~~~~~~~~~
  \label{eq:eqE2}
\end{eqnarray}
 Eq.\ref{eq:eqE1} and Eq.\ref{eq:eqE2} are solved by taking Laplace transform both with respect to time and space (see equations  Eq.\ref{eq:eqAA26} and Eq.\ref{eq:eqAA27} in Appendix C). Using Eq.\ref{eq:eqC24}, and by  putting $\nu_r=0$ in  Eq.\ref{eq:eqAA26} and Eq.\ref{eq:eqAA27},  after a series  of calculations we arrive at the following simple expression for length autocorrelation function:

\begin{eqnarray}
\mathcal{C}_{x}(t)=\frac{(v_s e^{-\nu_c^{\max}t}-v_g e^{-\frac{v_s}{v_g}\nu_c^{\max}t})}{v_s-v_g}  ~(\nu_r=0).
 \label{eq:eqE3}
\end{eqnarray}

In particular, in the limit $v_s\rightarrow v_g$ the above expression becomes,
\begin{eqnarray}
 \mathcal{C}_{x}(t)=e^{-\nu_c t}(1+ \nu_c^{\max} t)~~~~(\nu_r=0,v_g=v_s).
  \label{eq:eqE4}
\end{eqnarray}

\subsubsection{ Case 2. Linear aging, $\nu_c(\tau)=\beta \tau$} 

Eq.\ref{eq:eqB1} and Eq.\ref{eq:eqB2} are solved by taking transforms,   with  $\nu_c(\tau)$ given by  Eq.\ref{eq:eqA4a} (see Appendix C). By putting $\nu_r=0$ in Eq.\ref{eq:eqAA32} and Eq.\ref{eq:eqAA33}, and using Eq.\ref{eq:eqC23},  we find the  Laplace transform of the unnormalized correlation as

\begin{eqnarray}
\mathcal{\tilde C}_{x}^{\prime}(s)=\frac{v_g^2}{s \beta}+\frac {v_s e^{a^2s^2}\erfc(as)}{s^2 e^{b^2s^2}\erfc(bs)}\bigg[v_g \sqrt{\frac{2}{\pi \beta}}- \frac{v_g^2s}{\beta v_s} e^{c^2s^2}\erfc(cs)\bigg]  \nonumber \\ -\frac{v_s v_g}{s^2}\sqrt{\frac{2}{\pi \beta}}+\frac{v_g^2}{\beta}\sqrt{\frac{\pi}{2\beta}} e^{a^2s^2}\erfc(as)e^{c^2s^2}\erfc(cs),\nonumber \\ 
\label{eq:eqE5}
\end{eqnarray}
where the constants $a,b,c$ are defined as 

\begin{equation}
a=\frac{1}{\sqrt{2\beta}};~~c=\frac {v_g}{v_s \sqrt{2\beta}};~~b=a+c. \nonumber
 \label{eq:eqE6}
\end{equation}

As consistency checks for the expression in Eq.\ref{eq:eqE5}, we note that 
\begin{equation}
 \underset{s \to \infty}{\lim} s\mathcal{\tilde C}_{x}^{\prime}(s)=\underset{t \to 0}{\lim}\langle x(0)x(t)\rangle\equiv \langle x^2 \rangle=\frac{v_g^2}{\lambda \nu_c^{\max}}, 
\label{eq:eqE7}
\end{equation}

while 
\begin{equation}
\underset{s \to 0}{\lim} s\mathcal{\tilde C}_{x}^{\prime}(s)=\underset{t \to \infty}{\lim}\langle x(0)x(t)\rangle\equiv \langle x\rangle^2=\frac{2v_g^2}{\pi \lambda \nu_c^{\max}}, 
\label{eq:eqE8}
\end{equation}
 both of which agree with direct calculations, starting from the distribution in Eq.\ref{eq:eqC18}. 

It turns out difficult to extract more information about the time dependence of the function $\mathcal{C}^{\prime}_x(t)$ from the transform in Eq.\ref{eq:eqE5}, as the latter does not have a standard form which can be explicitly inverted. Therefore, to understand it better, we turn to numerical simulations; the results are presented in the next subsection.

Lastly,  the velocity autocorrelation ( numerical results will be discussed in the next subsection) is studied by defining a state variable $S(t)$, which takes value $+1$ or $-1$ depending on whether the filament is in growing or shrinking state, respectively, at time $t$. Then the instantaneous velocity of the filament is 

\begin{equation}
v(t)=v_g\bigg(\frac{1+S(t)}{2}\bigg)-v_s\bigg(\frac{1-S(t)}{2}\bigg)
\label{eq:eqC25}
\end{equation}
and the variable $S(t)$ switches from $+1$ to $-1$ at a (age-dependent) rate $\nu_c(\tau)$, and the reverse transition occurs at rate $\nu_r$. Therefore, the velocity autocorrelation can be easily expressed in terms of the state autocorrelation functions, defined as 
\begin{eqnarray}
\mathcal{C^{\prime}}_S(t)=\sum\limits_{i,j}\langle S(0)S(t)\rangle_{ij},~~~\mathcal{C}_S(t)=\frac{\mathcal{C^{\prime}}_S(t)-\langle S\rangle^2}{\sigma^2_S},~~
\label{eq:eqC26}
\end{eqnarray}
where $\sigma^2_S=1-\langle S\rangle^2$ is the variance in $S(t)$. In particular, when $v_s=v_g$, $v(t)=v_gS(t)$, and hence the normalized velocity autocorrelation $\mathcal{C}_v(t)=\mathcal{C}_S(t)$ itself.

\subsection{Numerical simulations}
\begin{figure}
\centering
\includegraphics[height=5.0cm,width=6.5cm]{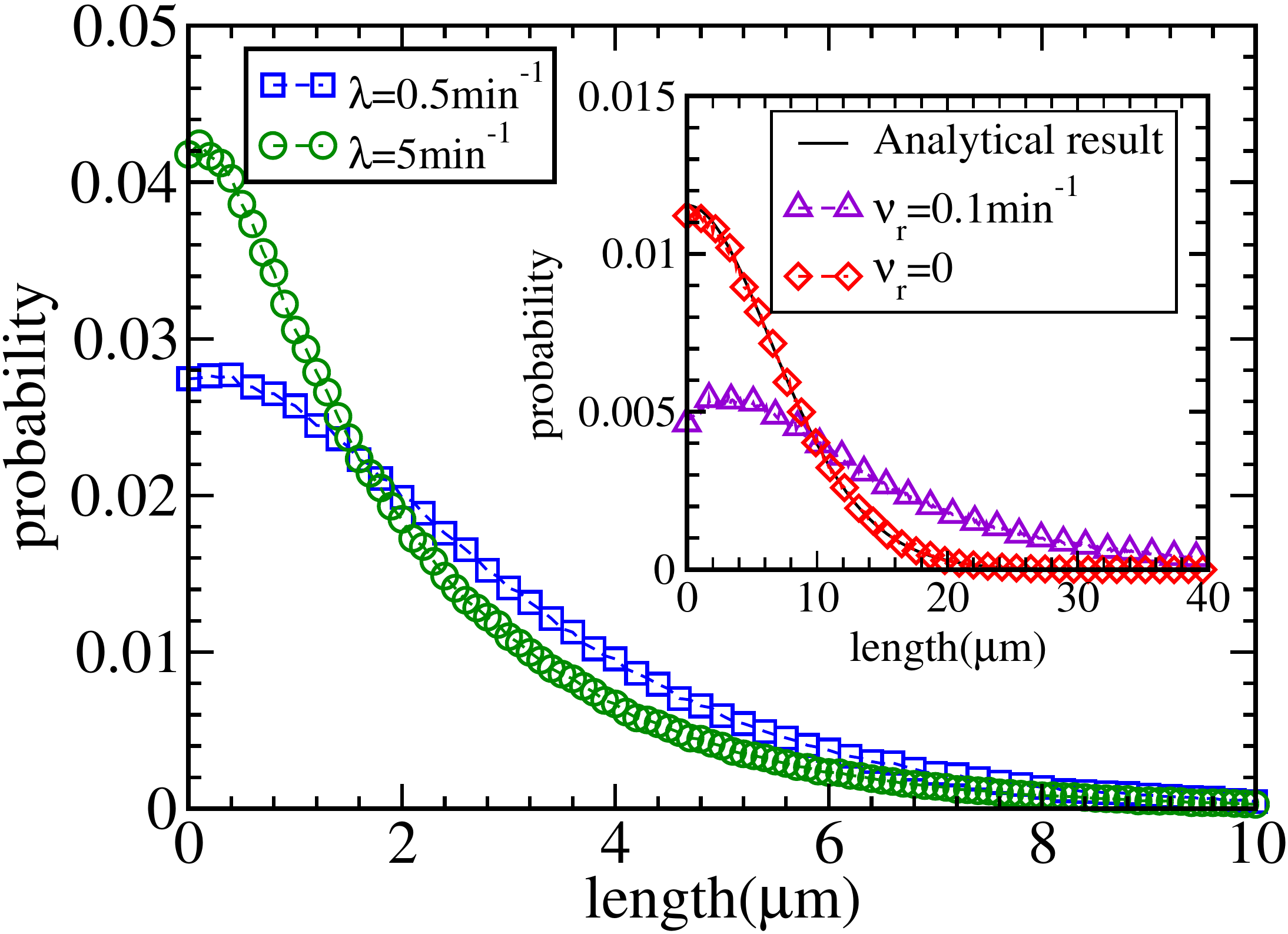}
 \caption { Microtubule  length distribution for the exponential model. The parameter values are $\nu=6\mathrm{min}^{-1}$, $\nu_r=0$, $\nu_c^{\mathrm {max}}=0.5 \mathrm{min}^{-1}$, $v_g=1\mu \mathrm{m~min^{-1}}$, ~$v_s=1.5\mu \mathrm{m~min^{-1}}$. In the inset, a comparison of length distribution for $\nu_r=0$ and $\nu_r=0.1\mathrm{min}^{-1}$ for $\lambda=0.05\mathrm{min}^{-1}$ is shown. The line is  shown for the analytical result given by Eq.\ref{eq:eqC12}. }
 \label{fig:fig3}
\end{figure}

\begin{figure}
\centering
\includegraphics[height=5.0cm,width=6.5cm]{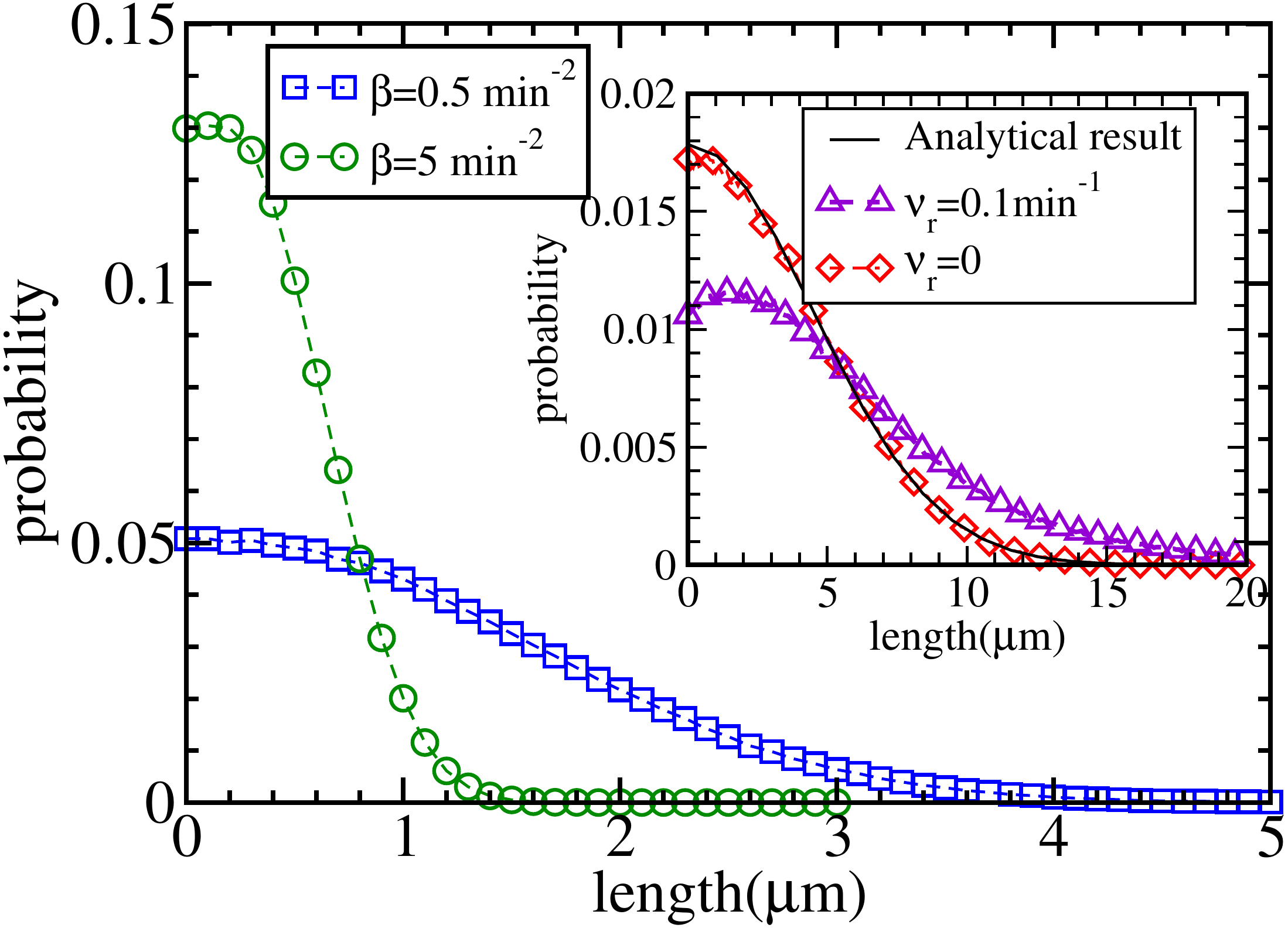}
 \caption { Microtubule  length distribution  for the linear model. The parameter values are $\nu=6 \mathrm{min}^{-1}$, $\nu_r=0$, $v_g=1\mu \mathrm{m~min^{-1}}$, ~$v_s=1.5\mu \mathrm{m~min^{-1}}$. In the inset, a comparison of length distribution for $\nu_r=0$ and $\nu_r=0.1\mathrm {min}^{-1}$ for $\beta=0.05\mathrm{min}^{-2}$ is shown. The line is  shown for the analytical result given by Eq.\ref{eq:eqC18}.}
 \label{fig:fig4}
\end{figure}

Given the difficulties in inverting the Laplace transforms, it became necessary to carry out stochastic numerical simulations to complete our study. Fixed time step (rather than Gillespie \cite{gillespie}) algorithm is used in simulations as the distribution of catastrophe time is non-exponential when aging is present. In the simulations, each of the microtubule is allowed to evolve  until a maximum time, independent of the others. We choose a time step $dt$ such that $0<Rdt<1$, where $R$ can be $\nu_c(\tau)$, $\nu_r$ and $\nu$.  The nucleation rate is chosen to be very large compared to the other rates in the simulation in order to achieve the boundary condition given by  Eq.\ref{eq:eqC4}.  A microtubule in the growing (shrinking)  state persists in that state and elongates by $v_g dt$ (shrinks by $v_s dt$) until it encounters catastrophe (rescue). 

We studied length distribution, average length as well as length and velocity autocorrelation functions.  Both exponential and linear catastrophes are studied. From the experiments reported in \cite{gardner}, for a tubulin concentration of $12\mu$M, we estimated that  $\lambda \approx 0.273 \mathrm {min^{-1}}$ and $\nu_c^{\max}\approx 0.3 \mathrm{min}^{-1}$, obtained by fitting the experimental data  to the function in  Eq.\ref{eq:eq1},  see Fig.\ref{fig:fig0}. Keeping these values as a reference point in the parameter space, in the simulations, we varied  $\lambda$ (keeping $\nu_c^{\max}$ fixed at $0.5\mathrm{min}^{-1}$) in the range  $0.05-5\mathrm {min}^{-1}$. For simulations with the linear aging model, we varied $\beta$ in  Eq.\ref{eq:eqA4a} in the range $0.05-5\mathrm {min}^{-2}$. Although rescue events were not observed in the experiments reported in \cite{gardner}, we considered the effects of non-zero rescue separately. The simulation results are discussed in detail now. \\

(i) {\it Length distribution and mean length}\\

The length distribution for exponential (Fig.\ref{fig:fig3}) and linear (Fig.\ref{fig:fig4}) aging models are shown, along with the corresponding analytical results. The main plots show the results with zero rescue; the effect of including small non-zero rescue frequency can be seen in the insets. 

\begin{figure}
\centering
\includegraphics[height=5cm,width=6.5cm]{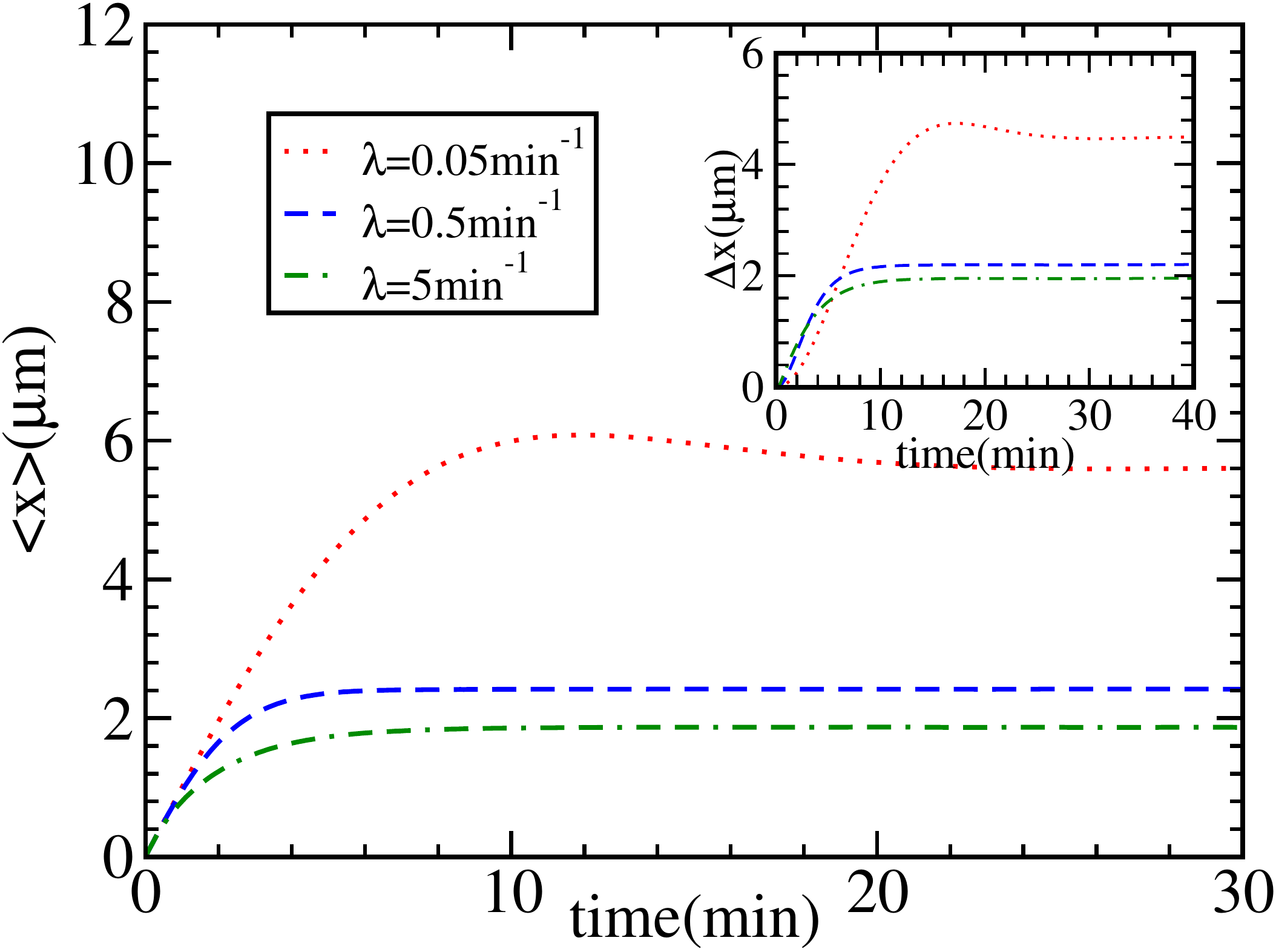}
 \caption { Time evolution of average microtubule  length   for the exponential  model, with the inset showing time evolution of standard deviation in length. The parameter values are $\nu=6\mathrm {min}^{-1}$, $\nu_r=0$, $\nu_c^{\max}=0.5 \mathrm {min}^{-1}$, $v_g=1\mu \mathrm{m~min^{-1}}$, ~$v_s=1.5\mu \mathrm{m~min^{-1}}$.}
 \label{fig:fig5}
\end{figure}

\begin{figure}
\centering

\includegraphics[height=5.0cm,width=6.5cm]{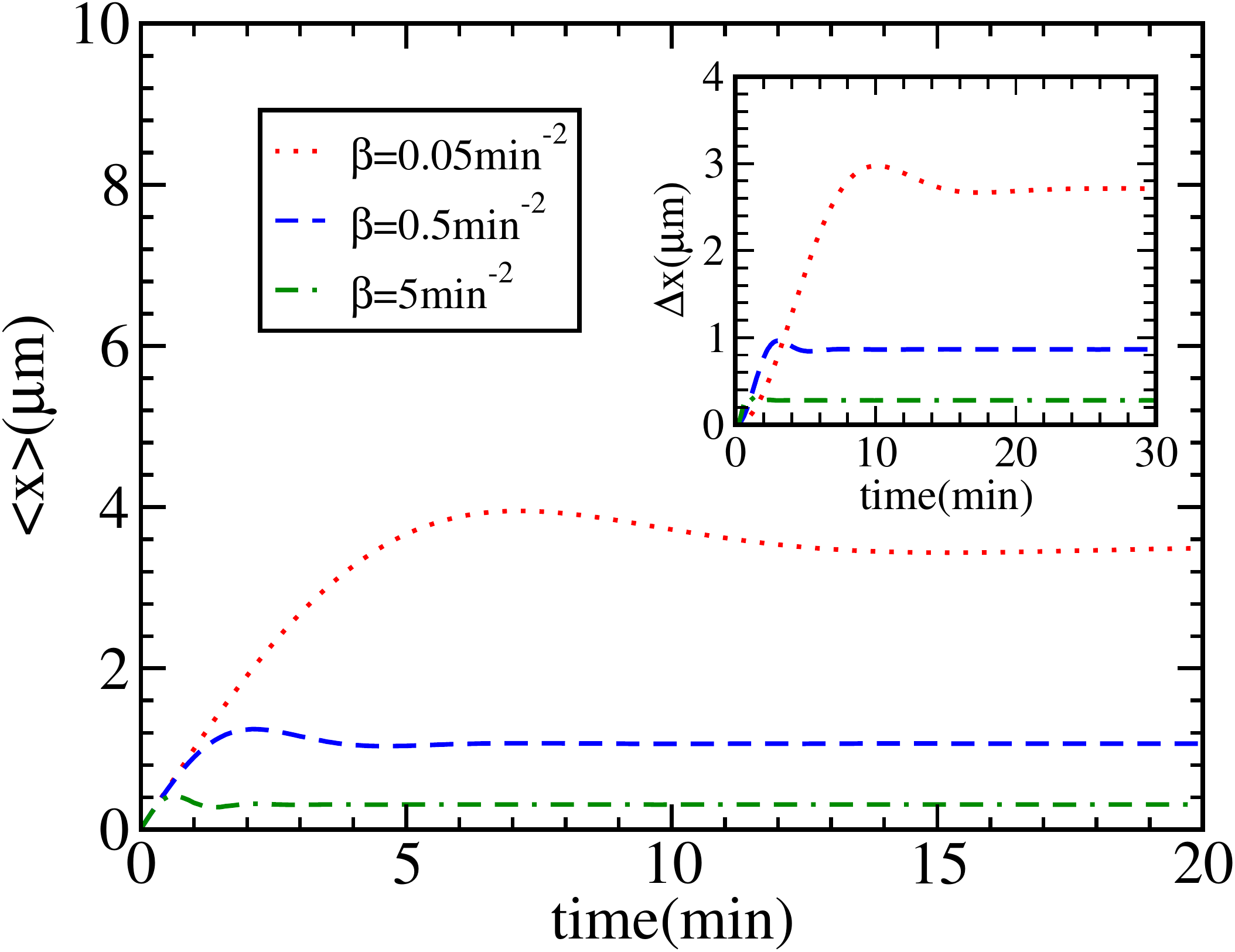}
 \caption { Time evolution of average  microtubule  length   for the linear model, with the inset showing time evolution of standard deviation in length. The parameter values are $\nu=6\mathrm {min}^{-1}$, $\nu_r=0$, $v_g=1\mu\mathrm{m~min^{-1}}$, ~$v_s=1.5\mu \mathrm{m~min^{-1}}$.}
 \label{fig:fig6}
\end{figure}

The time evolution of the average length is shown in Fig.\ref{fig:fig5} (exponential) and Fig.\ref{fig:fig6} (linear). As expected intuitively, the average length increases with the increase in the ``memory" of catastrophe (smaller $\lambda$ or $\beta$). It is observed from these plots that aging results in non-monotonic time evolution of average length for sufficiently small $\lambda$ (or $\beta$, in the linear model), indicating the possibility of oscillations. For the linear case, the peak in the curve remained noticeable even for large values of $\beta$ (Fig.\ref{fig:fig6}). 

\vspace{0.5cm}

(ii) {\it Length autocorrelation}\\

\begin{figure}
\includegraphics[height=5.0cm,width=6.5cm]{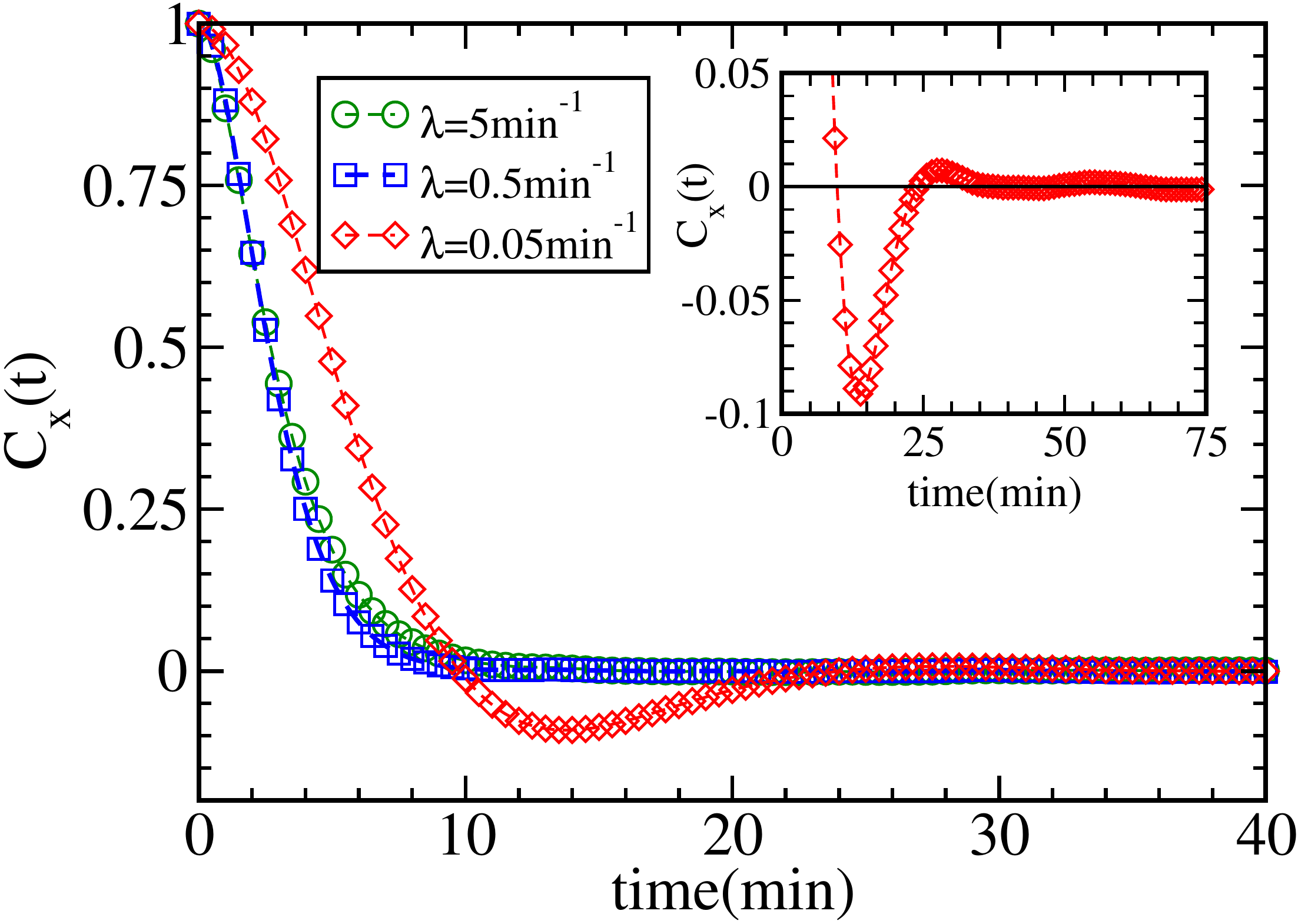}
 \caption{ Microtubule length autocorrelation function  for the exponential model. The parameter values are $\nu=6 \mathrm {min}^{-1}$, $\nu_r=0$, $\nu_c^{\max}=0.5 \mathrm {min}^{-1}$, $v_g=1\mu\mathrm {m~min}^{-1}$, ~$v_s=1.5\mu \mathrm {m~min}^{-1}$. In the inset, oscillatory behavior of the autocorrelation function corresponding to the parameter value $\lambda=0.05\mathrm{min}^{-1}$ is highlighted.}
 \label{fig:fig7}
\end{figure}

\begin{figure}
\centering
\includegraphics[height=5.0cm,width=6.5cm]{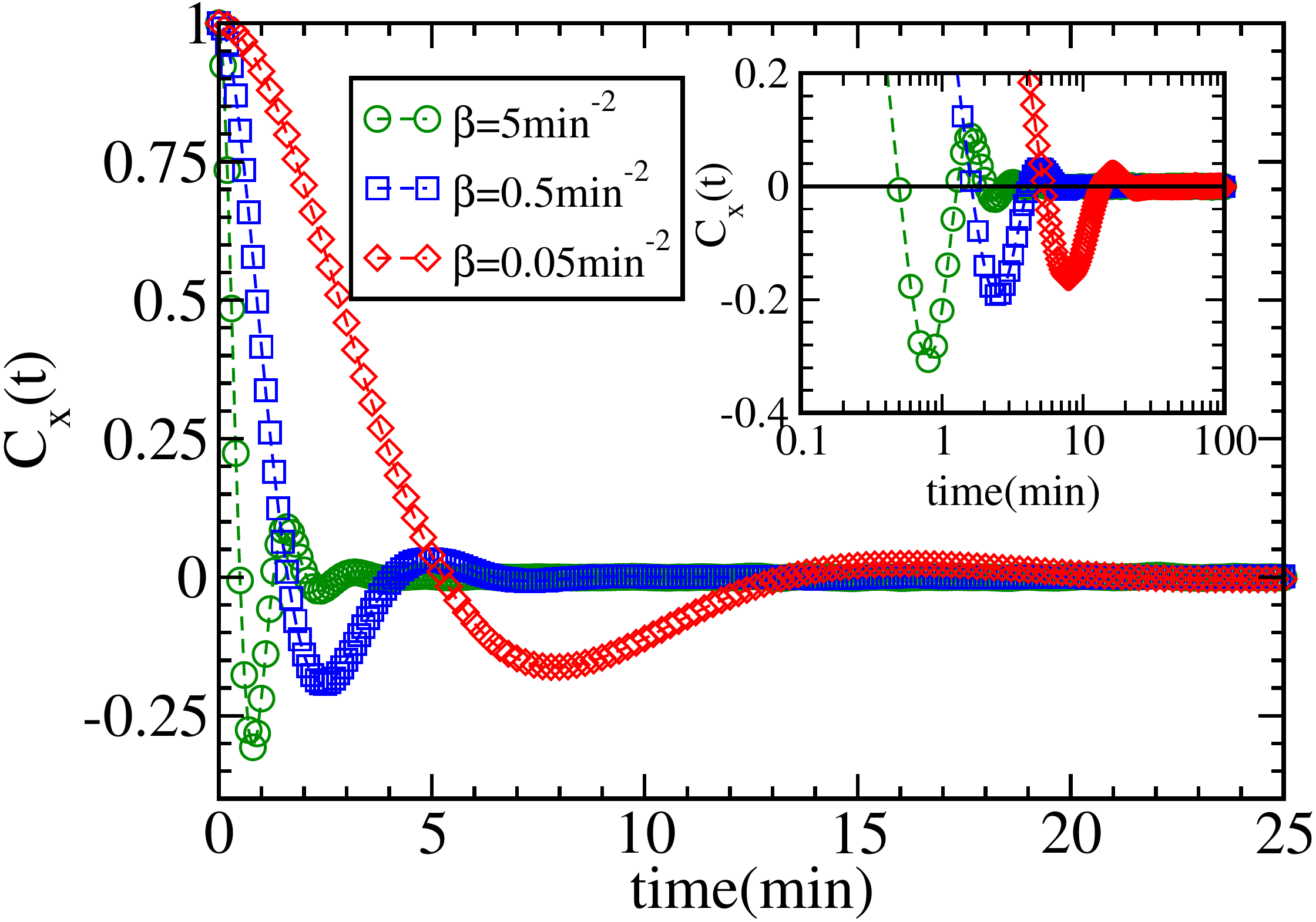}
 \caption{ Microtubule length autocorrelation function for the linear model. The parameter values are $\nu=6\mathrm {min}^{-1}$, $\nu_r=0$, $v_g=1\mu \mathrm {m~min}^{-1}$, ~$v_s=1.5\mu \mathrm {m~min}^{-1}$. In the inset, oscillatory parts of the autocorrelation functions are shown enlarged.}
 \label{fig:fig8}
\end{figure}

\begin{figure}
\centering
\includegraphics[height=5.3cm,keepaspectratio=true]{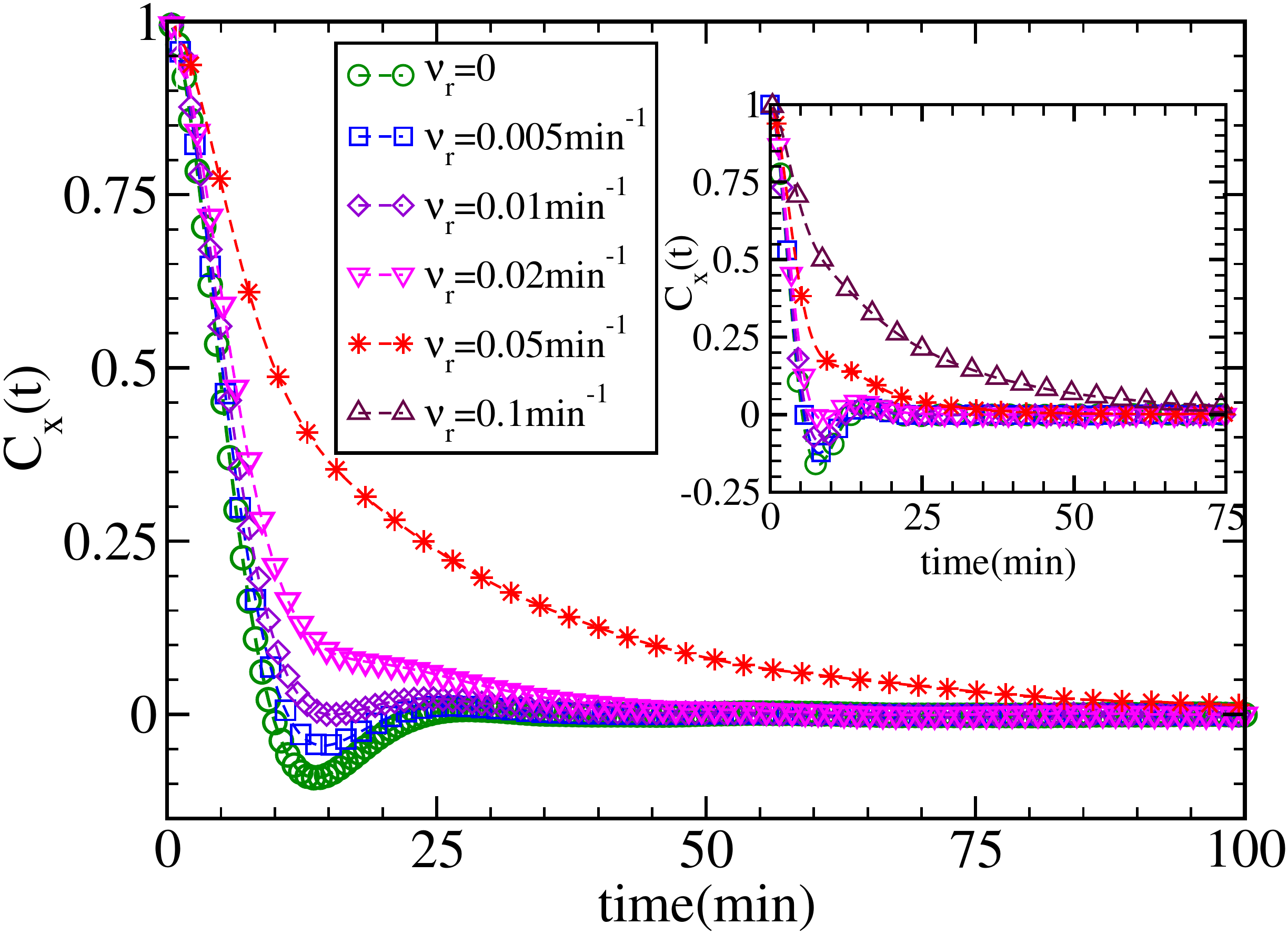}
 \caption{ Effect of rescue frequency on length autocorrelation function.  The main plot is shown for exponential model ($\lambda=0.05\mathrm {min}^{-1}$, $\nu_c^{\max}=0.5\mathrm {min}^{-1}$) and inset for linear model($\beta=0.05 \mathrm{min}^{-2}$). The parameter values are $\nu=6 \mathrm {min}^{-1}$, $v_g=1\mu\mathrm {m~min}^{-1}$, ~$v_s=1.5\mu \mathrm {m~min}^{-1}$. }
 \label{fig:fig9}
\end{figure}

\begin{figure}
\vspace{0.2cm}
\centering
\includegraphics[height=4cm,width=8.5cm]{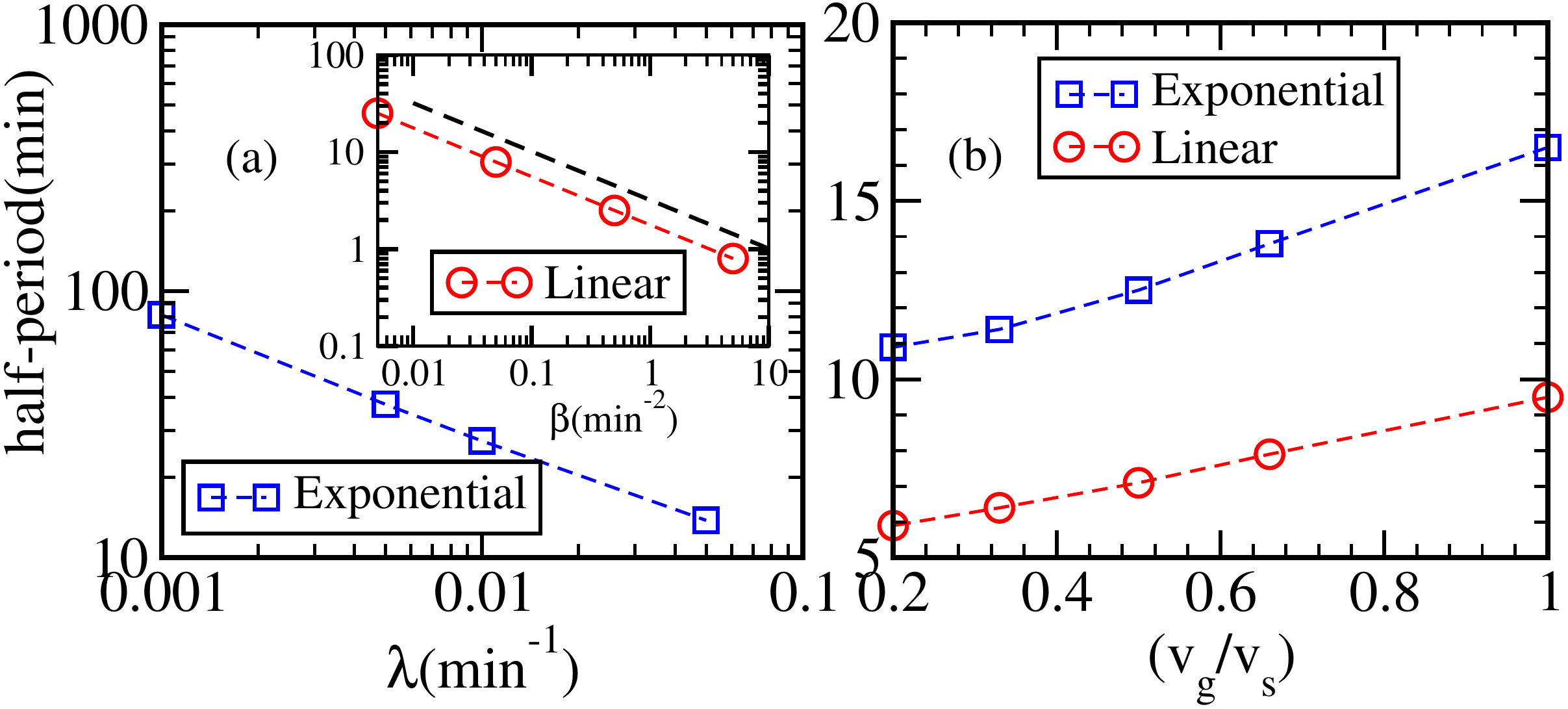}
 \caption{ Half period of length autocorrelation function. $(a)$ corresponds to half-period  as a function of $\lambda$, with inset showing the same for the linear model as a function of $\beta$. The parameter values are $\nu=6\mathrm {min}^{-1}$,  $v_g=1\mu \mathrm{m~min^{-1}}$, ~$v_s=1.5\mu\mathrm{m~min^{-1}}$ and for the exponential model $\nu_c^{\max}=0.5 \mathrm {min}^{-1}$. The dashed line is a fit of $\beta^{-\frac{1}{2}}$(Eq.\ref{eq:E13}). $(b)$ corresponds to half-period  as a function of $ v_g / v_s$. The parameter values are $\nu=6\mathrm {min}^{-1}$, $\nu_c^{\max}=0.5 \mathrm {min}^{-1}$,  $\lambda=0.05 \mathrm {min}^{-1}$ (exponential model) and $\beta=0.05 \mathrm {min}^{-2}$(linear model).}
\label{fig:fig10}
\end{figure}

\begin{figure}
\centering
\includegraphics[height=4.1cm,width=8.5cm]{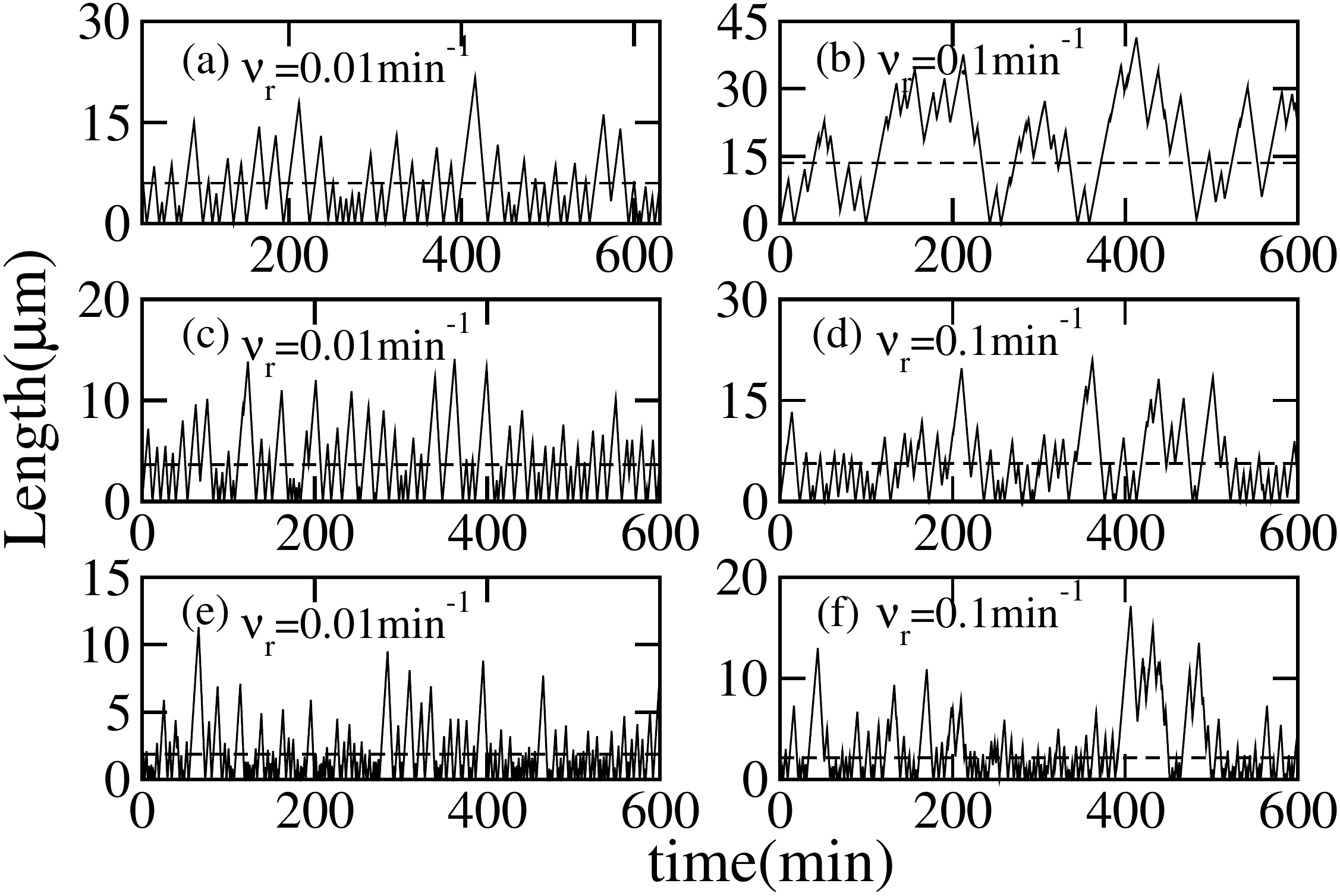}
 \caption{ Microtubule trajectories from stochastic simulation, with the dashed horizontal line denoting average length in the respective models studied.  $(a)$ and $(b)$ are for the exponential model with $\lambda=0.05 \mathrm{min^{-1}}$ and   $\nu_c^{\max}=0.5 \mathrm{min^{-1}}$. $(c)$ and $(d)$ are for the linear model with $\beta=0.05 \mathrm{min^{-2}}$. $(e)$ and  $(f)$ for the $\lambda\rightarrow\infty$ case with $\nu_c^{\max}=0.5\mathrm{min^{-1}}$. All the trajectories are simulated by fixing $v_g=1\mu \mathrm{m~min^{-1}}$, ~$v_s=1.5\mu\mathrm{m~min^{-1}}$. Only $a$ and $c$ correspond to oscillatory dynamics, as characterized by the negative lobe in the autocorrelation function. }
 \label{fig:fig11}
\end{figure}

Simulation results show that, in the presence of aging in catastrophe, the length autocorrelation has a pronounced (negative) minimum, followed by much weaker higher order extrema, characteristic of damped oscillations; see Fig.\ref{fig:fig7} and Fig.\ref{fig:fig8}. For the exponential model (see Fig.\ref{fig:fig7}), the minimum loses depth and eventually disappears with increasing $\lambda$ (eg., for $\nu_c^{\max}=0.5$min$^{-1}$, almost no trace of oscillations are seen for $\lambda> 0.5$min$^{-1}$). For the linear model, however, oscillations in the autocorrelation are seen to be present for a wide range of $\beta$ values studied, spanning three orders of magnitude.  

When rescue is switched on, the autocorrelation function loses its oscillatory character gradually in both models. This is shown in Fig.\ref{fig:fig9} for   exponential (main plot) and linear models (inset) respectively. In the exponential model, given the parameters $v_g=1\mu\mathrm {m~min}^{-1}$, ~$v_s=1.5\mu \mathrm {m~min}^{-1}$,  $\nu_c^{\max}=0.5\mathrm {min}^{-1}$ and  $\lambda=0.05 \mathrm {min}^{-1}$, the autocorrelation correlation appears to follow multi-exponential decay (without the negative lobe) when the rescue frequency $\nu_r$ is nearly $0.05\mathrm {min}^{-1}$; in the linear model, this occurs at $\nu_r=0.1\mathrm {min}^{-1}$, given $\beta=0.05\mathrm{min}^{-2}$. 

We also estimated the half-period of the oscillations in length (about its mean value) from the autocorrelation function, as the position of the first minimum in $C_x(t)$, determined from simulation results. Fig.\ref{fig:fig10} $(a)$ shows the dependence of the half-period on the aging rates $\lambda$ and $\beta$(inset): the logarithmic plots reveal that  the half-period $\propto \beta^{-\frac{1}{2}}$ in the linear model, which is expected. Based on the expression for autocorrelation function in Eq.\ref{eq:eqE5},  we propose the following scaling law for the linear model, when $\nu_r=0$: 

\begin{equation}
T_o=\frac{1}{\sqrt{\beta}}f\bigg(\frac{v_g}{v_s}\bigg),~~~~\mathrm{(linear~model)}
\label{eq:E13}
\end{equation}
where $T_o$ is the half-period of oscillations. It is clear that $f(x)$ is a increasing function of $x$; as $v_s\to\infty$, the microtubule will crash to the origin as soon as it undergoes catastrophe, and the period is determined by average time spent in the growing state, specified by Eq.\ref{eq:eqA6}. As $v_s$ is reduced, this time interval 
increases, which adds to the oscillation period. This is confirmed in simulations. Fig.\ref{fig:fig10}$(b)$ shows the half-period plotted as a function of $v_g/v_s$ in both exponential and linear models, which shows that the half-period (for fixed $\lambda$ and $\beta$) is an increasing function of $v_g/v_s$. 

For the exponential model, scaling considerations suggest that

\begin{equation}
T_o=\frac{1}{\sqrt{\beta}}g\bigg(\frac{\lambda}{\nu_c^{\mathrm{max}}},\frac{v_g}{v_s}\bigg),~~~\mathrm{(exponential~model)}
\label{eq:E13+}
\end{equation}
such that $\lim_{x\to 0}g(x,y)=f(y)$ for consistency with Eq.\ref{eq:E13}. Oscillations are present only when $\lambda/\nu_c^{\max}$ is sufficiently small. 
A few sample trajectories of a microtubule undergoing dynamic instability, under the age-dependent catastrophe are shown in Fig.\ref{fig:fig11}. 
\\

(iii) {\it Velocity autocorrelation}\\

\begin{figure}
\centering
\includegraphics[height=5cm,width=6.5cm]{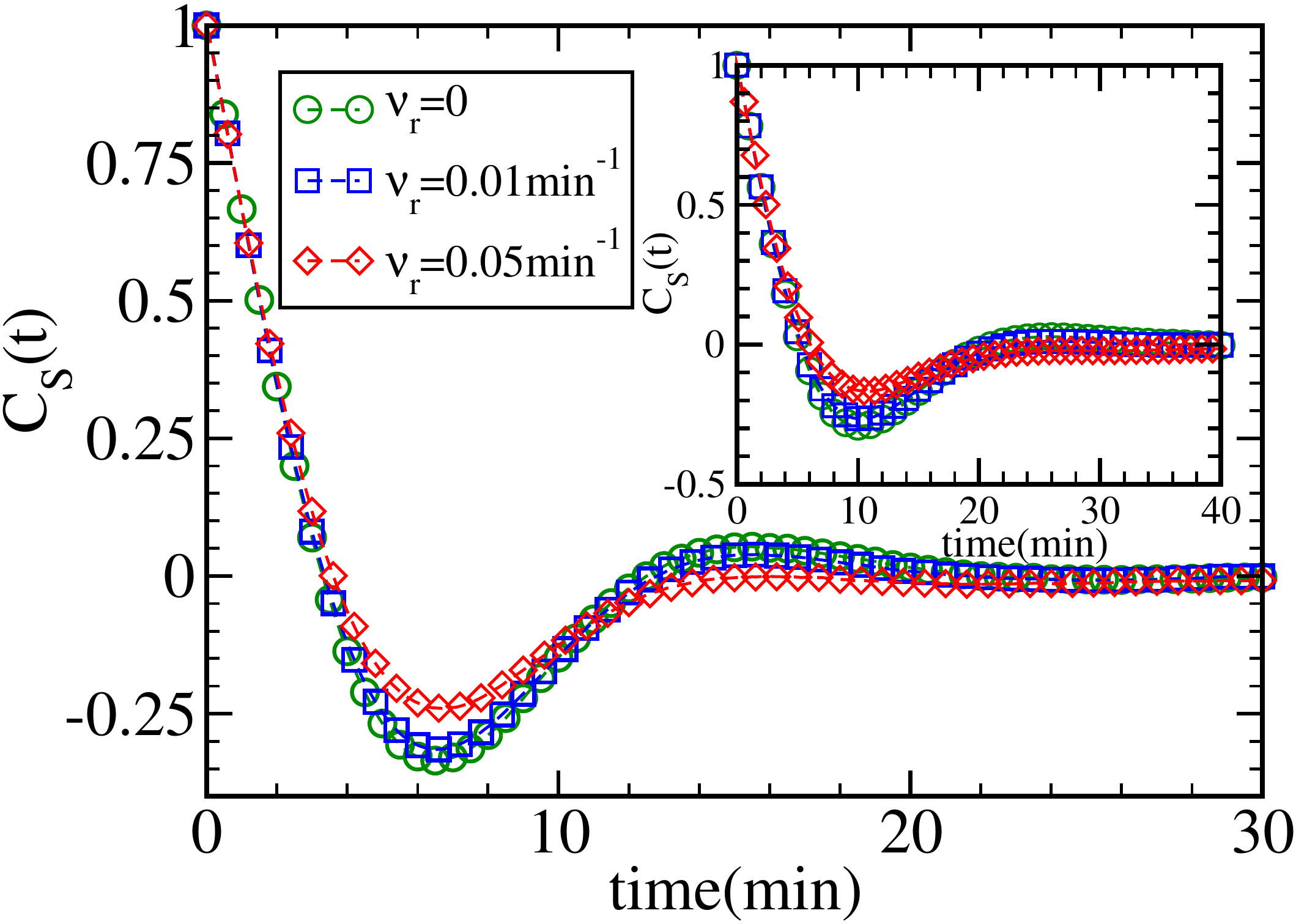}
\caption {State autocorrelation function $C_S(t)$  against time $t$ with a boundary at $x=0$. The main plot is shown for the linear model with $\beta=0.05\mathrm{min^{-2}}$  and inset for exponential model with $\lambda=0.05 \mathrm{min^{-1}}$. The other  parameter values are  $\nu=6  \mathrm{min}^{-1}$, $\nu_c^{\max}=0.5 \mathrm{min}^{-1}$, $v_g=1\mu \mathrm{m~min^{-1}}$, ~$v_s=1\mu \mathrm{m~min^{-1}}$.}
 \label{fig:fig12}
\end{figure}

The normalized velocity autocorrelation function, computed from the simulations, also shows oscillatory behavior (Fig.\ref{fig:fig12}) for sufficiently small values of $\lambda$ (and all values of $\beta$ studied in this paper). Also, increasing $\nu_r$ lifts up the negative lobe, weakening the oscillations. Interestingly, a comparison with Fig.\ref{fig:fig9} reveals that, for the same value of $\lambda$, the threshold rescue frequency for disappearance of oscillations is higher for velocity autocorrelation when compared to length, implying that for a certain range of parameter values, direction reversals are more regularly spaced in time when compared to excursions of length above and below the mean. 

Our results for velocity/state correlation are similar to the results reported by  Odde and Buettner \cite{odde1}. In this paper, the authors modeled dynamic instability as a two state jump process, but switching from one state to another state takes place in a sequence of steps such that residence times in both the states are characterized by (identical) gamma-distributions. Qualitatively, the effects of age-dependent (studied here) and multi-step (studied in Ref.\cite{odde1}) catastrophes are similar; however, in our model, only catastrophe is assumed to be age-dependent while rescue is treated as a first order process. 

\section{Summary and Conclusions}

Motivated by recent experimental observations on ``aging" in microtubule catastrophe, in this paper,  we have formulated a mathematical model to  investigate how the statistical properties of a microtubule population gets modified by it. Aging is characterized by a parameter $\lambda$, with units of inverse time, which we refer to as the aging rate in our model. Our results show that aging affects statistical properties of a microtubule population in important ways. The steady state length distribution is no longer a simple exponential decay as in the constant catastrophe case, while the condition for its existence is altered. The length and velocity autocorrelation functions develop a negative lobe for sufficiently small aging and rescue rates, which signifies oscillatory dynamics in the population. We characterized the oscillations using analytical calculations and scaling arguments, and checked their consistency with numerical simulations. Velocity (or growth state) oscillations in microtubule dynamics 
were studied earlier by Odde and Buettner \cite{odde1} using a model with {\it identical} gamma-distributed residence times in each state (i.e, both catastrophe and rescue were assumed to undergo aging in a similar way). Our study is similar in spirit, but with the following differences: (i) two different forms of aging are used for catastrophe, while rescue is treated as a first order process (in the absence of evidence to the contrary) (ii) we study how the steady state length distribution is modified in the presence of aging (iii) the length autocorrelation function is studied in detail, and we have derived an explicit mathematical expression for the same (in the linear aging model) in Laplace space.  

Microtubule length oscillations are likely to be relevant in the context of the well-known phenomenon of chromosome oscillations, although the latter is certainly a more complex process involving many other proteins and regulatory mechanisms. Nevertheless, we feel that it is interesting to make a broad comparison of our results with experimental observations on chromosome oscillations. In eukaryotic cells, soon after the onset of mitosis, each chromosome pair exists in a ``mono-oriented" state for a while, i.e., pulled/pushed only from one side  of the  mitotic spindle  \cite{salmon,west,kapoor} by microtubules attached to it through macromolecular structures called kinetochores.  A successful and efficient mitosis requires all the chromosomes to be ``bi-oriented" (i.e., bound to microtubule bundles emanating from the opposite poles simultaneously) at the spindle equator before segregation. It is known from experimental observations that both mono-oriented and bi-oriented chromosomes undergo regular 
oscillatory motion. 

Skibbens et al.  \cite{salmon} have showed that mono-oriented chromosomes in Newt lung cells exhibit non-sinusoidal oscillatory motion with a time period of $200-300$s. Similar oscillations were also observed in bi-oriented chromosomes, with a smaller period.  In PtK1 cells, the to and fro motion in mono-oriented chromosomes were observed to last for $\sim 1200 \mathrm{s}$  \cite{kapoor}, but the observed motion does not show the same degree of regularity as in the previous example. In HeLa cells, the autocorrelation function of translational movement of sister-kinetochores was experimentally measured more recently; the half-period of oscillations, measured as the position of the first minimum was found to be in the range $\sim 10 \mathrm{s}-40 \mathrm{s}$  \cite{amaro,amaro1}. With the available information, it is unclear whether aging and related effects in microtubule catastrophe plays a role in these phenomena;  however, given the fact that the observed period of oscillations of mono-oriented chromosomes 
\cite{salmon,kapoor} is close to the what was found in our simulations (Fig.\ref{fig:fig10}), we feel that further investigations in this direction could be worthwhile.\\ 

\acknowledgments

V.J thanks Erwin Frey for useful discussions during a visit to LMU, Munich. The authors acknowledge P.G. Senapathy Centre for Computing Resources, IIT Madras for computational support. The authors would also like to thank Dr. M.K. Gardner for sharing the relevant experimental data reported in \cite{gardner}.

\appendix
\section {PDFs in transformed space}
On taking Laplace transform of  Eq.\ref{eq:eqC2} and Eq.\ref{eq:eqC3} with respect to $x$, we  get
\begin{eqnarray}
\frac{\partial \tilde P_1^{\prime}(p,\tau)}{\partial \tau}+[v_g p+\nu_c(\tau)] \tilde P_1^{\prime}(p,\tau)=\delta(\tau) [\nu_r \tilde P_0(p)+J_0], \nonumber \\  
\label{eq:eqAA1}
\end{eqnarray}
\begin{eqnarray}
\tilde P_0(p)=\frac{J_0-\Phi(p)}{v_sp-\nu_r},~~~~~~~~~~~~~~~
\label{eq:eqAA2}
\end{eqnarray}
where $\Phi(p)=\int \limits_{0}^{\infty} \nu_c(\tau) \tilde P_1^{\prime}(p,\tau) d\tau$.
Substituting for $\tilde P_0(p)$ from Eq.\ref{eq:eqAA2} in Eq.\ref{eq:eqAA1},
\begin{eqnarray}
\frac{\partial \tilde P_1^{\prime}(p,\tau)}{\partial \tau}+[v_g p+\nu_c(\tau)]\tilde P_1^{\prime}(p,\tau)=~~~~~~~~~~~~~~~~~~~~  \nonumber \\  \delta(\tau)\bigg[ \frac{J_0 v_s p -\nu_r \Phi(p)}{v_sp-\nu_r} \bigg].~~~~~~~~~~~~~~
\label{eq:eqAA3}
\end{eqnarray}

Eq.\ref{eq:eqAA3} is a first order non-homogeneous partial differential equation in $\tau$. We assume a solution of the form,
\begin{eqnarray}
\tilde P_1^{\prime}(p,\tau)=h(p,\tau) \exp \bigg( -v_g p \tau -\int \limits_{0}^{\tau} \nu_c(\tau') d\tau' \bigg).
\label{eq:eqAA4}
\end{eqnarray}

After substituting Eq.\ref{eq:eqAA4} in  Eq.\ref{eq:eqAA3} we get,
\begin{eqnarray}
h(p,\tau)= \frac{J_0 v_s p -\nu_r \Phi(p)}{v_sp-\nu_r}.
\label{eq:eqAA5}
\end{eqnarray}

Therefore the total solution to Eq.\ref{eq:eqAA3} is given by 
\begin{eqnarray}
 \tilde P_1^{\prime}(p,\tau)= \bigg[\frac{J_0 v_s p -\nu_r \Phi(p)}{v_sp-\nu_r}\bigg]\times~~~~~~~~~~~~~~~ \nonumber \\ \exp \bigg( -v_g p \tau -\int \limits_{0}^{\tau} 
 \nu_c(\tau') d\tau'\bigg).
 \label{eq:eqAA6}
\end{eqnarray}
Using the above equation we get,  

\begin{eqnarray}
\tilde \Phi(p)=\frac{  J_0 v_s p \zeta(p)  }{v_sp-\nu_r (1-\zeta(p))},~
 \label{eq:eqAA7}
\end{eqnarray}
where 
\begin{eqnarray}
 \zeta(p)= \int \limits_{0}^{\infty} \nu_c(\tau) \exp \bigg( -v_g p \tau -\int \limits_{0}^{\tau} \nu_c(\tau') d\tau' \bigg) d\tau.~~
 \label{eq:eqAA8}
\end{eqnarray}
The function $\eta(p)$, which appears in Eq.\ref{eq:eqC7} of the main text is defined as,
\begin{eqnarray}
 \eta(p)=\int \limits_{0}^{\infty} \exp\bigg(-v_g p \tau -\int \limits_{0}^{\tau} \nu_c(\tau') d\tau'\bigg) d\tau.  
 \label{eq:eqAA11}
\end{eqnarray} 
As mentioned in the main text, all further analysis is done by putting $\nu_r=0$. In this case, Eq.\ref{eq:eqC7} and Eq.\ref{eq:eqC8} in main text become, 
\begin{eqnarray}
\tilde P_1(p)=J_0\eta(p)
\label{eq:eqAA14}
\end{eqnarray}
and
\begin{eqnarray}
\tilde P_0(p)= J_0\frac{ 1-\zeta(p)}{v_sp} .
\label{eq:eqAA15}
\end{eqnarray}

\section{Length distribution} 
\subsubsection{Exponential aging}
In this case, $\zeta(p)$ and $\eta(p)$ are evaluated from the integrals given by Eq.\ref{eq:eqAA8} and  Eq.\ref{eq:eqAA11} respectively, with $\nu_c(\tau)$ substituted from Eq.\ref{eq:eq1}.

\begin{eqnarray}
\zeta(p)=\frac{\nu_c^{\max}}{v_gp+\nu_c^{\max}}- \frac{v_gp}{v_gp+\nu_c^{\max}} e^{\alpha}\times ~~~~~~~~~~~~\nonumber \\ \alpha^{-\frac{v_g p+\nu_c^{\max}}{\lambda}}\gamma\bigg(\frac{v_g p+\nu_c^{\max}}{\lambda}+1,\alpha\bigg), ~~~~
\label{eq:eqAA16}
\end{eqnarray} 
\begin{eqnarray}
\eta(p)=\frac{1}{\lambda}e^{\alpha}\alpha^{-\frac{v_gp+\nu_c^{\max}}{\lambda}} \gamma\bigg(\frac{v_gp+\nu_c^{\max}}{\lambda},\alpha\bigg).~~~~~~~~
 \label{eq:eqAA17}
\end{eqnarray}
Using the recurrence relation \cite{abramowitz}, 
\begin{eqnarray}
 \gamma(a+1,b)= a \gamma(a,b)- b^a e^{-b}, 
  \label{eq:eqAA18}
\end{eqnarray} 
we get a relation between $\zeta(p)$ and $\eta(p)$ as given below,
\begin{eqnarray}
\zeta(p)=1-v_g p ~\eta(p).
\label{eq:eqAA19}
\end{eqnarray}
Therefore, Eq.\ref{eq:eqAA15} can be rewritten as
\begin{eqnarray}
 \tilde P_0(p)=  \frac{ v_g}{v_s} J_0 \eta(p).
 \label{eq:eqAA21}
\end{eqnarray}
Using the inverse Laplace transform, 
\begin{eqnarray}
 \mathcal{L}^{-1}[a^{-p}\gamma(p,a)]= e^{-ae^{-x}}, 
 \label{eq:eqAA22}
\end{eqnarray}
we find the Laplace inverse of  Eq.\ref{eq:eqAA14} and Eq.\ref{eq:eqAA21} with $\eta(p)$ substituted from  Eq.\ref{eq:eqAA17}.  The resulting expression for the total PDF is given by Eq.\ref{eq:eqC12} in the main text.
\subsubsection{Linear aging}
The integrals given by Eq.\ref{eq:eqAA8} and Eq.\ref{eq:eqAA11} are evaluated using Eq.\ref{eq:eqA4a}. The expressions for $\zeta(p)$ and $\eta(p)$
are, respectively, given as, 

\begin{eqnarray}
\zeta(p)=1-\sqrt{\frac{\pi}{2\beta}}v_gp ~e^{\frac {v_g^2 p^2}{2\beta}}\erfc{\bigg[\frac{v_gp}{\sqrt{2\beta}}\bigg]}, 
\label{eq:eqAA23}
\end{eqnarray}

\begin{eqnarray}
\eta(p)=\sqrt{\frac{\pi}{2\beta}} ~e^{\frac {v_g^2 p^2}{2\beta}}\erfc{\bigg[\frac{v_gp}{\sqrt{2\beta}}\bigg]}. ~~~
\label{eq:eqAA24}
\end{eqnarray}
From Eq.\ref{eq:eqAA23} and Eq.\ref{eq:eqAA24}, it can be seen that Eq.\ref{eq:eqAA19} holds true in the linear model as well. 
A useful inverse Laplace transform to invert Eq.\ref{eq:eqAA14} and Eq.\ref{eq:eqAA15} in this case is given as \cite{abramowitz}, 
\begin{equation}
 \mathcal{L}^{-1}\bigg[e^{a^2p^2}\erfc[ap]\bigg]=\frac{e^{-\frac{x^2}{4a^2}}}{a\sqrt{\pi}}
 \label{eq:eqAA25}
\end{equation}
 and the final expression for the $P(x)$ is given by Eq.\ref{eq:eqC18} in the main text. 
\\
\section{Length autocorrelation}
\subsubsection{ Constant catastrophe: the limit $\lambda\rightarrow\infty$}
Conditional probabilities obtained from  Eq.\ref{eq:eqE1} and Eq.\ref{eq:eqE2} after taking Laplace transform with respect to space and time,  are given by
\begin{widetext}
\begin{eqnarray}
G_{1j}(p,s|x_0)=\frac{e^{-px_0}[(s-v_sp)\delta_{1j}+\nu_r] +v_s(s-v_sp)G_{0j}(x=0,s|x_0) }{(s-v_sp+\nu_r)(s+v_gp+\nu_c^{\max})-\nu_c^{\max}\nu_r},
\label{eq:eqAA26}
\end{eqnarray}
\begin{eqnarray}
G_{0j}(p,s|x_0)=\frac{e^{-px_0}[(s+v_gp)\delta_{0j}+\nu_c^{\max}]-v_s(s+v_gp)G_{0j}(x=0,s|x_0) }{(s-v_sp+\nu_r)(s+v_gp+\nu_c^{\max})-\nu_c^{\max}\nu_r}.
\label{eq:eqAA27}
\end{eqnarray}
\end{widetext}
We use the convergence property of the conditional PDFs to find $G_{0j}(x=0,s|x_0)$. Thus, at the boundary, expressions for the Green's functions are given by
\begin{eqnarray}
G_{0j}(x=0,s|x_0) =\frac{(s+v_g \beta_s) \delta_{0j}+ \nu_c^{\max}}{v_s(s+v_g\beta_s)} e^{-\beta_s x_0},~~~~\nonumber \\
\label{eq:eqAA28}
\end{eqnarray}
where $\beta_s=-\frac{A(s)}{2}+\sqrt{\frac{A(s)^2}{4}+B(s)}$, with 
\begin{eqnarray}
 A(s)=\frac{s(v_s-v_g)+v_s\nu_c^{\max}-\nu_rv_g}{v_gv_s},~~~~~~~~\\
 B(s)=\frac{s(s+\nu_c^{\max}+\nu_r)}{v_gv_s}.~~~~~~~~~~~~~~
\end{eqnarray}
In the limit $\nu_r=0$, Eq.\ref{eq:eqAA28} becomes,  
\begin{eqnarray}
G_{0j}(x=0,s|x_0) =\frac{s(v_s+v_g) \delta_{0j}+v_s \nu_c^{\max}}{s v_s (v_s+v_g)} e^{-\frac{s}{v_s}x_0}.\nonumber \\
\label{eq:eqAA29}
\end{eqnarray}
In the steady state, the PDFs take the simple form,
 \begin{eqnarray}
 P_1(x)=\frac{J_0}{v_g}  e^{-\frac{\nu_c^{\max} x}{v_g}},             
 \label{eq:eqAA30}
\end{eqnarray}
\begin{eqnarray}
 P_0(x)=\frac{J_0}{v_s}  e^{-\frac{\nu_c^{\max} x}{v_g}},
 \label{eq:eqAA31}
\end{eqnarray}
where $J_0=\nu_c^{\max} v_s/(v_g+v_s).$

\subsubsection{Linear aging}

Solutions of Eq.\ref{eq:eqB1} and Eq.\ref{eq:eqB2} for the linear model in transformed space are respectively given as,  
\begin{widetext}
 \begin{eqnarray}
G_{1j}(p,s|x_0)=\frac{e^{-px_0}[(s-v_sp)\delta_{1j}+\nu_r] +v_s(s-v_sp)G_{0j}(x=0,s) }{[s-v_sp+\nu_r(1-\zeta(p,s)]}\eta(p,s),~~~~~
\label{eq:eqAA32}
\end{eqnarray}
\begin{eqnarray}
 G_{0j}(p,s|x_0)=\frac{e^{-px_0}[\delta_{0j}- \zeta(p,s)( \delta_{0j}-1)] -v_s (1-\zeta(p,s))G_{0j}(x=0,s) }{[s-v_sp+\nu_r(1-\zeta(p,s)]},~~~~~
 \label{eq:eqAA33}
\end{eqnarray}
\end{widetext}
with 
\begin{eqnarray}
\zeta(p,s)=1-\sqrt{\frac{\pi}{2\beta}}(s+v_gp) ~e^{\frac {(s+v_g p)^2}{2\beta}}\erfc{\bigg[\frac{s+v_gp}{\sqrt{2\beta}}\bigg]} \nonumber \\
\label{eq:eqAA34}
\end{eqnarray}
and 
\begin{eqnarray}
\eta(p,s)=\sqrt{\frac{\pi}{2\beta}}e^{\frac {(s+v_g p)^2}{2\beta}}\erfc{\bigg[\frac{s+v_gp}{\sqrt{2\beta}}\bigg]}. ~~
\label{eq:eqAA35}
\end{eqnarray}
Using similar procedure mentioned above, we fix $G_0j(x=0,s|x_0)$ in the limit $\nu_r=0$ as
\begin{eqnarray}
 G_{0j}(x=0,s|x_0)=\frac{e^{-\frac{sx_0}{v_s}}[v_s- (1-\delta_{0j})s(vs+vg) \psi(s)]}{sv_s(v_s+v_g)\psi(s)},\nonumber \\
 \label{eq:eqAA36}
\end{eqnarray}
 where 
\begin{eqnarray}
\psi(s)= \sqrt{\frac{\pi}{2\beta}} ~\exp{\bigg(\frac{(v_s+v_g)^2 s^2}{2v_s^2\beta}\bigg)}\erfc\bigg[{\frac{(v_s+v_g) s}{v_s\sqrt{2\beta}}}\bigg].~~~~~~~~
\label{eq:eqAA37}
\end{eqnarray}
Finally, the steady state length distribution required to evaluate  Eq.\ref{eq:eqC23} is given by Eq.\ref{eq:eqC18} and  the expression for unnormalized correlation in the transformed space is given by Eq.\ref{eq:eqE5}. 
\end{document}